\documentclass[vecphys]{svmult}

\usepackage{makeidx}         
\usepackage{graphicx}        
\usepackage{multicol}        
\usepackage{cite}            
\usepackage[bottom]{footmisc}
\usepackage{bm}
\usepackage{amsmath}
\usepackage{amssymb}
\usepackage{color}

\makeindex             

\begin{document}
\setcounter{chapter}{9}
\title{All-Dielectric Nanophotonic Structures:\\ Exploring the Magnetic Component of Light}
\titlerunning{All-dielectric Nanophotonic Structures}
\author{B.~Hopkins* \and A.~E.~Miroshnichenko \and Y.~S.~Kivshar}
\authorrunning{B.~Hopkins \and A.~E.~Miroshnichenko \and Y.~S.~Kivshar}
\institute{Nonlinear Physics Centre, Australian National University, Canberra 2601, Australia \quad *\texttt{ben.hopkins@anu.edu.au}}

\maketitle

\begin{abstract}
We discuss nanophotonic structures composed of high-index dielectric nanoparticles and present several basic approaches for numerical study of their collective optical response. 
We also provide comparison on the collective optical properties of dielectric and plasmonic structures, and review experimental demonstrations of Fano resonances in all-dielectric nanoparticle oligomers.
\end{abstract}

\noindent

\section{Introduction}

Recent progress in nanophotonics 
\index{Nanophotonics}
is often associated with the study of resonant plasmonic structures that localize light at the nanoscale or provide control over far-field scattering. 
However, a new branch of nanophotonics emerged through the manipulation of optically-induced Mie-type electric and magnetic resonances in dielectric and semiconductor nanoparticles with high refractive index.
\index{Dielectric nanoparticles}
High-index resonant nanoparticles offer advantages over their plasmonic counterparts in terms of reduced dissipative losses and access to resonant enhancement of both the electric {\sl and} the magnetic near-field\cite{Evlyukhin2010, Krasnok2011_Jetp, GarciaEtxarri2011}.
\index{Optical magnetism}
The coexistence of electric and magnetic resonances has also enabled effective realization of Kerker's conditions\cite{Kerker1975} for reflectionless scattering 
\index{Directional scattering}
from individual dielectric nanoparticles\cite{Nieto2011, Rolly2012, Geffrin2012,  Kuznetsov2012,Evlyukhin2012NL}, a development that subsequently allows various new phenomena for manipulation of directional light scattering~\cite{Liu2012_ACSNANO,Fu2013, Liu2014_CPB} and optical nanoantenna applications~\cite{Krasnok2012_OE,Filonov2012,Krasnok204_nanoscale}.
This Chapter summarizes several basic numerical approaches applied to the analysis of all-dielectric nanophotonic structures, while also comparing our results with those of plasmonic structures, and their functionalities.
Within this scope, we place emphasis on one particular area that has garnered significant attention in recent years: the study of Fano resonances
\index{Fano resonance}
 in nanoparticle oligomers and cluster structures~\cite{Miroshnichenko2010, Lukyanchuk2010}.
 \index{Oligomers}
The aforementioned directionality of single dielectric nanoparticles can be considered as an effect of interference in the scattering along particular directions, while Fano resonances correspond to a resonant interference in the {\sl total} scattering of an object. 
These two phenomena are then interdependent once a dielectric nanoparticle oligomer can exhibit a Fano resonance. 
In this regard, we take the opportunity to discuss the formation, more generally, of Fano resonances  in nanoparticle oligomer geometries.   
Fano interference has been predominantly described for specific plasmonic oligomers, where a directly-excited, super-radiant mode interferes destructively with an indirectly-excited ``dark" (or trapped) mode. 
The indirect-excitation of the dark mode is generally attributed to a near-field interaction, the so-called {\it hybridization} of plasmons, which provides a coupling channel between the different resonant subsystems of a plasmonic nanostructure.  
This description resembles a classical oscillator model~\cite{Joe2006, Gallinet2011}, where a driven oscillator is able to resonantly couple its external driving force into an adjacent oscillator.  
Yet it is less clear how to apply this description when the separation of a collective system into distinct resonant subsystems is ambiguous, or simply arbitrary.  
Further, Fano resonances have since been predicted in all-dielectric symmetric oligomers~\cite{Miroshnichenko2012}, despite the absence of plasmon hybridization between nanoparticles.
The formal treatment presented herein, considers Fano resonances instead from the resonances of the collective system, {\it i.e.} without separation into subsystems.  
This approach is able to present a single common model for Fano resonances in both plasmonic and all-dielectric nanoparticle scattering systems.  
Specifically, Fano resonances can be rigorously attributed to fact that the eigenmodes of lossy resonating systems are nonorthogonal, and therefore: {\sl they can interfere with each other}.
This simple model also sheds light on why Fano resonances occur for a wide range of nanoparticle systems, including highly symmetric nanoparticle oligomers, without requiring any additional complexity be added to interactions between nanoparticles.  

To this end, the outline of our Chapter is as follows.  
We present first the approach used for modeling dielectric nanoparticle systems in Section~\ref{sec:1},  before then discussing, in Section~\ref{sec:2}, the use of eigenmodes to describe collective resonances, including the formation of Fano resonances.
We then conclude by reviewing some recent experimental developments in producing Fano resonances with dielectric nanoparticle oligomers in Section~\ref{sec:3}.  

\section{Modeling All-Dielectric Nanoparticle Systems}
\label{sec:1}
\subsection{Magnetism in Nanophotonics}
\begin{figure}[h!]
\centerline{\includegraphics[width=\textwidth]{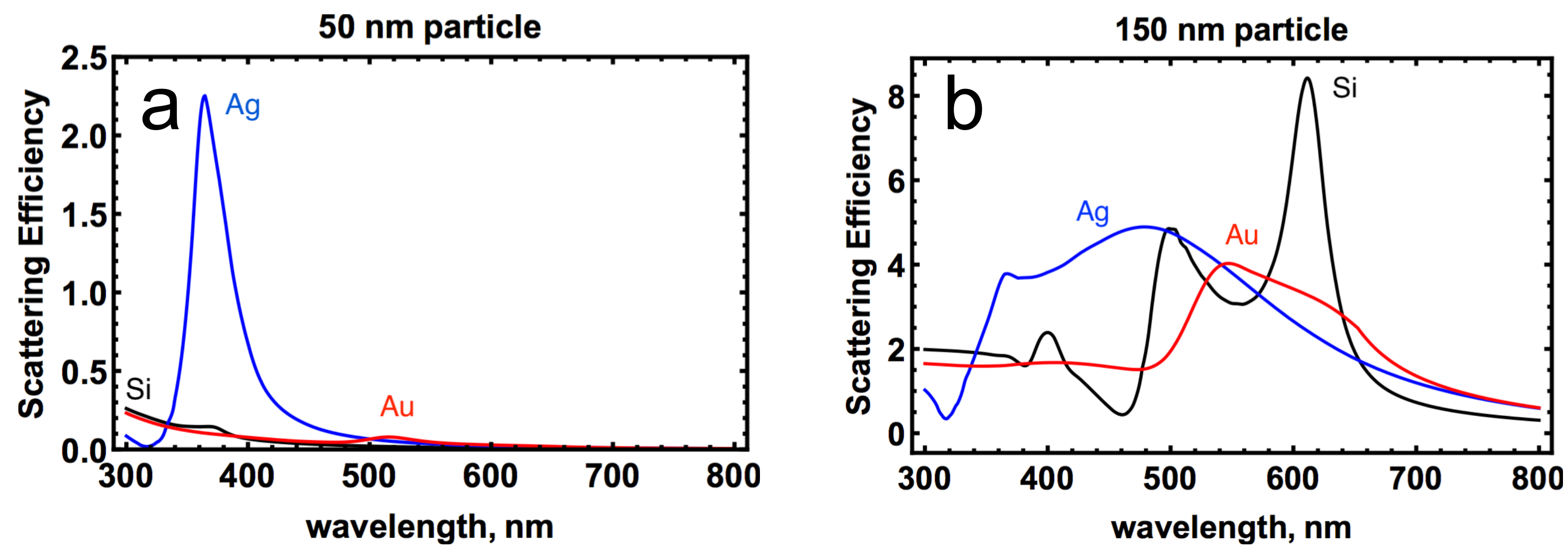}}
\caption{Comparison of the scattering cross-sections of gold, silver and silicon nanoparticles
\index{Dielectric nanoparticles} 
of the same geometrical sizes: (a) diameter 50~nm, and (b) diameter 150~nm. These results demonstrate that in addition to a new type of resonances dielectric particles scatter light more efficiently than plasmonic ones for larger sizes.  Produced after A.~I.~Kuznetsov\cite{KuznetsovSPIE2015}.}
\label{fig:comparison}
\end{figure}

Physically, electromagnetism describes retarded forces between charges.
Electric and magnetic fields are defined as heralds of linear and rotational force on charges, a basis that has been historically necessary, or at least convenient, because it corresponds to the way natural materials respond.
Indeed, atoms and other neutral compositions of charged matter dominantly behave as per electric and magnetic dipoles: oscillating charges and circulating currents. 
Material then arises when defining regions of closely packed electric and magnetic dipoles with homogenized volumes of electric and magnetic dipole densities, two densities which we refer to as {\it permittivity} and {\it permeability}. Homogenization, when defined as such, inherently corresponds to material as found in nature. 
However, this macroscopic description need not align with the advent of metamaterials, nanoantennas, and complex structured media generally~\cite{Sheinfux2014}. 
Macroscopic assemblies of subwavelength {\it meta-atoms} will not necessarily respond to fields as per linear or rotational movement of charge; the constituent point sources in such a homogenized material are not necessarily electric or magnetic dipoles. 
Nor is there any guarantee that resonant field distributions of a meta-atom will remain stable field oscillations at frequencies detuned from that of their resonance. 
Indeed, even compact arrangements of coupled dipolar nanoparticles can collectively exhibit such higher-order optical behavior on a subwavelength scale~\cite{HopkinsPoddubny2013}.
Therefore, underlying the rapid development of complex nanophotonic structures is a need to re-evaluate how we understand the optical properties of complex structured objects and media.
The development of all-dielectric nanostructures is one such area demanding attention, because they inherently operate near the limit of point dipole analogies: typical resonant nanostructures made of a high-index dielectric material utilize field retardation, and hence {optical sizes} on the order of the operating wavelength, to access Mie-like resonances.  
The most apparent example of this being simple high-index dielectric nanoparticles that are implicitly able to exhibit {\sl both} electric and magnetic dipolar resonances.
\index{Magnetic resonance}
These have similar magnitude to the electric dipole resonances of plasmonic nanoparticles, as is seen in Figure~\ref{fig:comparison}.
\index{Mie resonance}
This figure also demonstrates how the low Ohmic losses in dielectric materials offers reduced damping of resonances, thereby providing higher quality factors when compared to plasmonic counterparts.  
Extending the electric and magnetic single particle response to arbitrary, high refractive index, dielectric materials also translates these resonances across a large spectrum, owing to the scaling properties of the resonances with refractive index and nanoparticle size~\cite{GarciaEtxarri2011}.

The seemingly small change of having both electric and magnetic dipole resonances can have dramatic effects on the scattering behavior of single nanostructures\cite{Kerker1975, Staude2013, Lukiyanchuk2015}, but also collective nanoparticle clusters\cite{Chong2014, HopkinsFilonov2015} and arrays\cite{Pfeiffer2013, Decker2015, Chong2015, Arbabi2015}.
In this Chapter, complex dielectric nanostructures are first discussed in terms of free currents and polarization currents, because these are the underlying physical sources of fields within any arbitrary system, and thereby encapsulate the complete optical properties.
We then present the dipole model as a practical simplification that allows direct investigation of the dominant resonances of compact nanoparticle systems.
This two-model approach is used for understanding the physics of nanoparticle cluster geometries consisting of high-index dielectric nanoparticles, and the simultaneous interplay of electric and magnetic fields.

\subsection{Radiation by Internal Current Distributions}
We begin our scattering analysis of linear optical systems by acknowledging that any arbitrary time-varying electric and magnetic fields can be described as a distribution of harmonic fields in a Fourier series representation.
Furthermore, for monochromatic response, there is no particular need to recognize the distinction between free current and polarization current, given conductivity $ {\bar{\sigma}}$ and susceptibility $ {\bar{\chi}}$ can be incorporated into an effective permittivity ${\bar{\epsilon}}$ that relates electric field $\mathbf{E}$ to a {\it total} electric current $\mathbf{J}$.
\begin{align}
{\bar{\epsilon}}\equiv ( {\bar{\chi}}+1)\epsilon_0 - \frac{ {\bar{\sigma}}} {i \omega}  \quad \Rightarrow \quad \mathbf{J} (\mathbf{r}) e^{-i\omega t} = - i \omega  \big( {\bar{\epsilon}}(\mathbf{r})- \epsilon_0 \big)  \mathbf{E}(\mathbf{r}) e^{-i\omega t} \label{eq:induced current}
\end{align}
There is also the simplification that most optical materials have a negligible difference in permeability compared to their respective background medium, allowing us to neglect the radiation from any magnetization current.
As such, the field $\mathbf{E_s}$  scattered from an arbitrary structure can be described by radiation from a distribution of electric current, and can be expressed in terms of dyadic Green's functions~\cite{Yaghjian1980}.
\index{Green's function}
\begin{align}
\mathbf{E_s}(\mathbf{r})  = i \omega \mu_0  \int _{V}  {\bar{G}_{0}}(\mathbf{r},\mathbf{r}')\mathbf{J}(\mathbf{r}') \;{{\mathrm{dr}'}^{3}}  \label{eq:scatt}
\end{align}
where $k$ is the wavenumber, $\omega$ is the angular frequency, $\epsilon_0$ and $\mu_0$ are the permittivity and permeability of the background medium, and  $V$ is the volume of the scattering object, which we assume here to be finite.
The remaining part of the expression is the free space dyadic Green's function  $\bar{G}_{0}$:
\begin{align}
 {\bar{G_0}}(\mathbf{r},\mathbf{r}')=  \mathop{\mathrm{P.V.}} \Big(\mathrm{\bar{I}}+ \frac{1}{k^2} \boldsymbol{\nabla \nabla}\Big)  &\frac{e^{i k R}}{4 \pi   R}- \mathrm{\bar{\,L}} \frac{\delta (\mathbf{r}-\mathbf{r}')}{k^2 }
\end{align}
where $R = \left| \mathbf{r}-\mathbf{r}'\right|$, the $\mathrm{P.V.}$ implies a principal value exclusion of $\mathbf{r}'=\mathbf{r}$ when performing the integration in  Equations~\ref{eq:induced current} and \ref{eq:current equation}, and $ \mathrm{\bar{\,L}}$ is the source dyadic necessary to account for the shape of this exclusion\cite{Yaghjian1980}.
Notably, we can then use this expression for the scattered field to find the current induced by an external electric field $\mathbf{E_0}$.
In particular, using Equation~\ref{eq:induced current}, the current can found from the total internal field  $\mathbf{E} = \mathbf{E_0}+\mathbf{E_s}$, producing the following relation between the induced field and the external field:
\begin{align}
 - i \omega \big({\bar \epsilon(\mathbf{r}) - \epsilon_0}\big) \,\mathbf{E}_{0}(\mathbf{r})   =&{\mathbf{J}(\mathbf{r})}  - \frac{k^2}{\epsilon_0} \big(  {\bar\epsilon(\mathbf{r})  - \epsilon_0} \big)  \int _{V}  {\bar{G_0}}(\mathbf{r},\mathbf{r}')\mathbf{J}(\mathbf{r}') \;{\mathrm{dr'}^3}  \label{eq:current equation}
\end{align}
As such, Equation~\ref{eq:current equation} and Equation~\ref{eq:scatt} will completely describe the currents induced and fields radiated, respectively, by any arbitrary finite structure in the absence of magnetization.
This is a comprehensive description of general scattering systems, and is therefore useful for finding broad analytical conclusions as a result of geometry.
Below, we will  use it to discuss the relation between far-field interference features and the nonorthogonality of eigenmodes for the induced current.
It is, however, highly nontrivial to obtain a general, excitation-independent, modal solution for Equation~\ref{eq:current equation} with even very simple geometries, plainly because the model inherently describes the complete dynamics of a given object's scattering, encapsulating all resonances, irrespective of their magnitude.
One existing numerical approach for calculating the general scattering properties of any arbitrary object from the resonances of surface current distributions is implemented through the {\it OpenModes} software, which deals with complexity by searching for only a subset of the complex frequency resonances\cite{Powell2014}.
Here, however, we will focus on arrangements of simple nanoparticles, so-called {\it oligomers}, 
 \index{Oligomers}
 which utilize the coupling between nanoparticles, rather than nanoparticle geometry, to produce complex optical features such as Fano resonances\cite{Hentschel2010, Chong2014} and near-field control\cite{HopkinsLiu2013, Rahmani2013}.
For these nanoparticle oligomers, we are able to restrict our analysis to the dominant optical properties of the collective geometry by only considering the dominant optical response of the individual nanoparticles.
In particular, the model that will be used extensively throughout the remainder of this Chapter is to consider only the dipole-order responses of individual nanoparticles.
As we will show, this approach suffers from only minor limitations in accuracy while offering dramatic simplifications in modal analysis.

\subsection{Dipole Models}
\begin{figure}
\centerline{\includegraphics[width=\textwidth]{{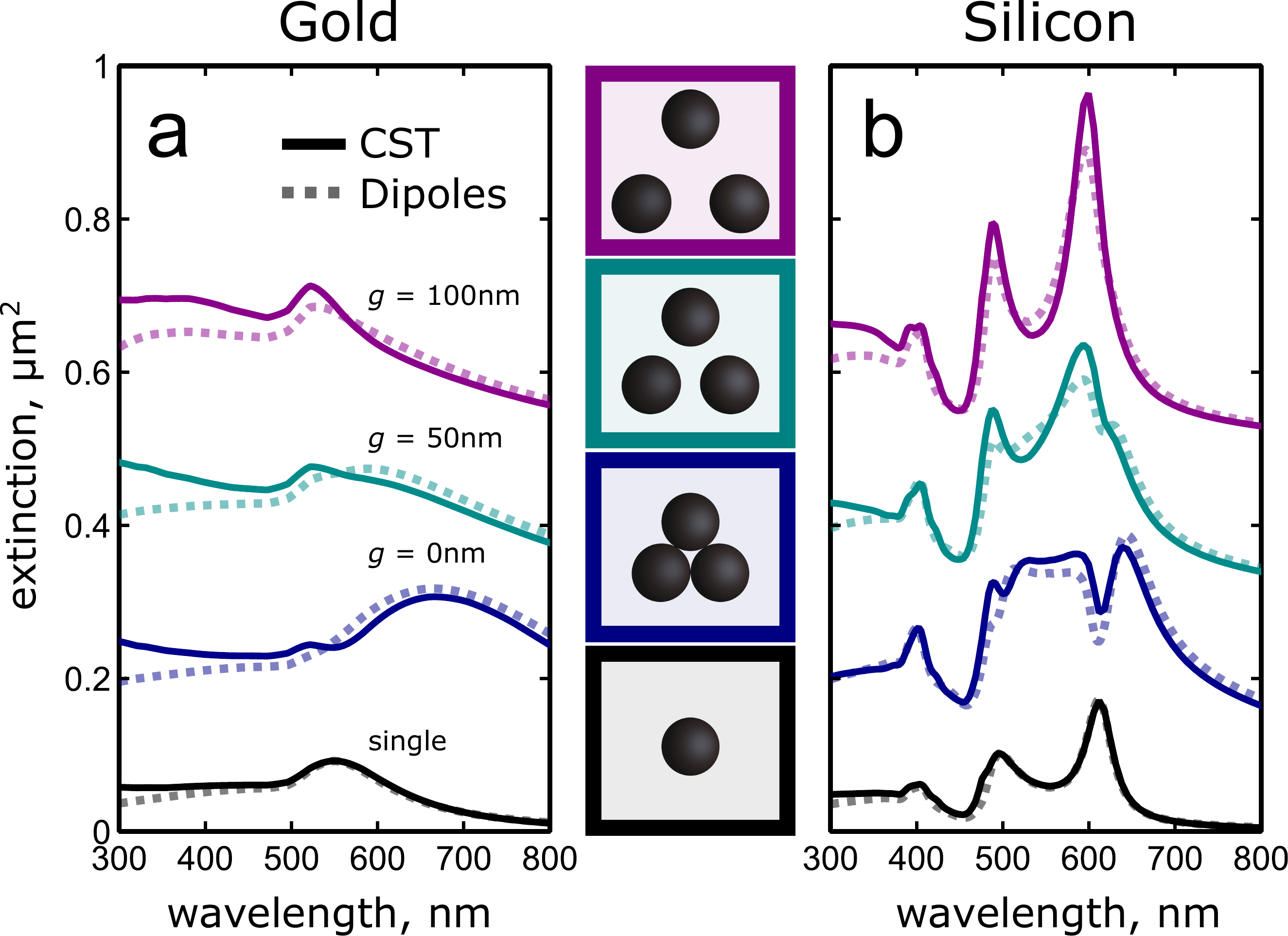}}}
\caption{A comparison of the extinction cross-section calculated using CST Microwave Studio with that calculated using the dipole model in Equation~\ref{eq:Dmodel}.
Calculations are for 150~nm nanospheres made of (a) gold and (b) silicon, when arranged as symmetric trimers with varying separation $g$ between nanoparticles.   }
\label{fig:trimerComparison}
\end{figure}

\index{Dipole model}
The constituent nanoparticles of oligomers are, in general, of simple geometry and subwavelength in size, which implies that their individual optical responses can be entirely described in terms of resonant dipole moments\cite{Mulholland1994, Evlyukhin2010}.
We will begin in the simpler case of plasmonic nanoparticles, where the individual nanoparticle response is dominantly an electric dipole, and we can therefore use the dipole approximation\cite{Draine1994} to describe the optical properties of the oligomer.
In particular, each nanoparticle's electric dipole moment $(\mathbf{p}_{i})$ is related to the externally-applied electric field $(\mathbf{E_0})$ as:
\begin{align}
\mathbf{p}_{i}  = \alpha_{{\scriptscriptstyle E}}^{(i)}\epsilon_{0}\mathbf{E_{0}}(\mathbf{r}_{i})&+\alpha_{\scriptscriptstyle E}^{(i)}k^{2}
\sum \limits_{j\neq i} \bar{G}_{0}(\mathbf{r}_{i},\mathbf{r}_{j}) \cdot \mathbf{p}_{j}
\label{eq:ED model}
\end{align}
Here, $\alpha_{{\scriptscriptstyle E},i}$ is the electric dipole polarizability of the $i^{\mathrm{th}}$ particle and we can write the effect of $\bar{G}_{0}$ more explicitly as:
\begin{petit}
\begin{align}
\bar{G}_{0}(\mathbf{r}',\mathbf{r}) \cdot \mathbf{p} = \frac{ e^{i k R }}{4 \pi  R}\bigg[ \Big(1 + \frac{i}{k R}  - \frac{1}{k^2 R^2} \Big)\, \mathbf{p} - \Big(1 + \frac{3 i}{k R}  - \frac{3}{k^2 R^2}\Big)( \mathbf{n}\cdot\mathbf{p})\,\mathbf{n}\bigg]  \label{eq:EEcoupling}
\end{align}
\end{petit}%
\index{Green's function}%
where $\mathbf{n}$ is the unit vector pointing from $\mathbf{r}$ to $\mathbf{r}'$.
The expression in Equation~\ref{eq:ED model} forms a matrix equation of rank $3N$, where $N$ is the number of dipoles, and it can be solved for an arbitrary excitation as per an ordinary matrix equation.
In Figure~\ref{fig:trimerComparison}a, we present the validity of this model for a symmetric trimer arrangement of gold nanospheres, where the electric dipole polarizabilities of the nanospheres are calculated through $a_1$ scattering coefficients from Mie theory\cite{Mie1908, Bohren1983}, {\it cf.} Equation~\ref{eq:polarizabilities}. 
\index{Mie resonance}
Even with very small gaps between the spheres, the dipole model offers an accurate prediction of the trimer's response, with the exception of that coming from the single particle electric quadrupole response.
However, the situation becomes notably different when we utilize high-index dielectric nanoparticles.  
\index{Dielectric nanoparticles}
Simple dielectric nanospheres will also exhibit a magnetic dipole resonance 
\index{Magnetic resonance} 
in addition to an electric dipole resonance.
Indeed, we can express the electric and magnetic dipole polarizabilities in terms of the $a_1$ and $b_1$ scattering coefficients, which are known from Mie theory for the case of spheres\cite{Mie1908, Bohren1983}.
\begin{align}
\alpha_{\scriptscriptstyle E} &= \frac{6 i \pi a_1}{k^3} \;,\qquad
\alpha_{\scriptscriptstyle H} = \frac{6 i \pi b_1}{k^3} \label{eq:polarizabilities}
\end{align}
To this end, the dipole model in Equation~\ref{eq:ED model} can be extended to include both electric and magnetic dipoles\cite{Mulholland1994}
\index{Optical magnetism}
\begin{subequations}\label{eq:Dmodel}
\begin{align}
\mathbf{p}^{(i)}  = \alpha_{E}^{(i)}\epsilon_{0}\mathbf{E_{0}}(\mathbf{r}_{i})&
+\alpha_{E}^{(i)}k^{2}
  \bigg(\underset{j\neq i}{{\sum}}\bar{G}_{0}(\mathbf{r}_{i},\mathbf{r}_{j})\cdot  \mathbf{p}^{(j)}
-\frac{1}{c_0}\,\nabla\times\bar{G}_{0}(\mathbf{r}_{i},\mathbf{r}_{j}) \cdot \mathbf{m}^{(j)}\bigg) \\
 \mathbf{m}^{(i)} =  \alpha_{H}^{(i)}\mathbf{H_{0}}(\mathbf{r}_{i})&
 +\alpha_{H}^{(i)}k^{2}  \bigg(\underset{j\neq i}{{\sum}}\bar{G}_{0}(\mathbf{r}_{i},\mathbf{r}_{j}) \cdot \mathbf{m}^{(j)}
+c_0\,\nabla\times\bar{G}_{0}(\mathbf{r}_{i},\mathbf{r}_{j}) \cdot \mathbf{p}^{(j)}\bigg)
\end{align}
\end{subequations}
where $\mathbf{p}^{(i)}$ ($\mathbf{m}_i$) is the electric (magnetic) dipole moment of the $i^{\mathrm{th}}$ particle, $\bar{G}_{0}(\mathbf{r}_{i},\mathbf{r}_{j}) $ is the free space dyadic Green's function between the $i^{\mathrm{th}}$ and $j^{\mathrm{th}}$ dipole, $\alpha_{E}^{(i)}$ ($\alpha_{H}^{(i)}$) is the electric (magnetic) dipole polarizability of the $i^{\mathrm{th}}$ particle, $c_0$ is the speed of light in free-space and $k$ is the free-space wavenumber.
The extra cross-coupling terms are given according to:
\begin{align}
\nabla\times\bar{G}_{0}(\mathbf{r}',\mathbf{r}) \cdot \mathbf{p}
&=\frac{ e^{i k R }}{4 \pi  R}\bigg(1 + \frac{i}{k R}  \bigg) \mathbf{n}\times\mathbf{p} \label{eq:EMcoupling}
\end{align}%
\index{Green's function}%
Notably, Equation~\ref{eq:ED model} is the discrete equivalent to the continuous current distribution model in Equation~\ref{eq:current equation}, and in Equation~\ref{eq:Dmodel} it is extended to account for magnetization currents in addition to polarization currents.
In other words, the magnetic dipoles introduced into Equation~\ref{eq:Dmodel} behave as per sources of magnetization current.
This description of circulating current in a nanoparticle is, at least macroscopically, indistinguishable to a source of true magnetization that derives from material permeability.
Indeed, this can be understood due to the fact that, regarding atoms in natural materials, permeability was originally defined to describe circulating currents.
This is analogous to circulating polarization current as per Equation~\ref{eq:induced current}, and a high-index dielectric nanoparticle remains smaller than the resonant wavelength, even if much larger than either atoms or molecules.
As such, dielectric nanoparticles allow a straightforward avenue to access effective material properties with both permittivity and permeability.
However, we previously mentioned that a key advantage of nanostructures is to go beyond conventional materials.
In this regard, we simply need to acknowledge that a single nanoparticle will rarely act in isolation: it radiates and couples with other nearby nanoparticles to form collective modes, and the individual nanoparticles no longer respond proportionally to the applied electric and magnetic fields as per permittivity and permeability.
In the next section we investigate such collective optical responses rigorously by presenting an eigenmode approach for describing the optical responses of coupled nanoparticle scattering systems.

\section{Eigenmodes of Nanoparticle Oligomers}
\label{sec:2}

\subsection{Resonances, Polarizability, and Eigenmodes}
One of the most important properties of nanoscale structures is their resonant response, and how one might utilize a given resonance's characteristics: quality factor, electric or magnetic field localization, directionality, chiral preferentiality, or any number of other foreseeably useful properties.
In this regard, the majority of quantitative approaches to characterize a resonance rely on a multipolar decomposition of a given resonance, either in terms of spherical harmonics (to obtain a spherical multipole decomposition), or Cartesian multipoles, and their various correction factors~\cite{Chen2011, Grahn2012, Miroshnichenko2015}.
These decompositions can then be used to, for instance, define polarizabilities or impedances for each multipolar dimension\cite{Arango2013}.
However, it worth recognizing that multipolar decompositions  are inherently dependent on the choice of origin when performing the decomposition, and that multipoles ultimately remain a non-unique {\sl basis} for the scattering responses.  
This can occasionally lead to a large number of multipoles being necessary to describe a single resonance, such as when either the multipole basis, or its choice of origin, is not necessarily representative of the given resonance.
Furthermore, while there is an intuitive choice of origin for simple nanostructures, it can be much less obvious for complex nanostructures or arrangements of nanoparticles, including nanoparticle oligomers.
These issues can also impact other analytical tools in nanophotonics, such as the $T$-matrix method, which rely on the same multipole decompositions.

In an attempt to understand the resonances of a system more fundamentally, and in a way that is independent of the choice of origin, we can instead consider the eigenmodes of the electric current system in the presence of a driving field.
Moreover, Equation~\ref{eq:current equation} has an associated eigenmode equation, where an eigenmode $|v \rangle$ has a  current distribution ${\mathbf{J}_v}$ and eigenvalue $\lambda_v$ that satisfies:
\begin{align}
i \omega   \lambda_v {\mathbf{J}_v}(\mathbf{r})    =&  - \big( {{\bar {\epsilon}}(\mathbf{r}) - \epsilon_0}\big)^{-1}{{\mathbf{J}_v}(\mathbf{r})} +\frac{1}{\epsilon_0}\int _{V}  {\bar{{G}}_0}(\mathbf{r},\mathbf{r}'){\mathbf{J}_v}(\mathbf{r}') \;{\mathrm{dr}'}^3   \label{eq:eigenmode equation}
\end{align}
These eigenvalues and eigenmodes represent an origin-independent basis of polarizabilities or impedances, and the associated stable current distributions, respectively.
Furthermore, the set of eigenmodes represents a basis for the optical response where each basis vector represents a current distribution that is subject to energy conservation in isolation.
In a mathematical sense (refer to Equation~\ref{eq:extinction definition}), the real component of the eigenvalue must be greater than zero to be passive, or must be less than zero to be active, where \textit{active} means inputting net energy into the system and {\it passive} simply refers to being not active.
The final characteristic to recognize is that the complex frequency where an eigenvalue becomes zero corresponds to a self-sustaining current distribution, otherwise recognized as a resonance of the system.
In fact, given the eigenmodes will almost always form a complete and linearly-independent basis, \textsl{every} such self-sustaining resonance must be associated with at least one zero eigenvalue.
In summary, the eigenmode decomposition obtains a complete set of underlying resonances at \textsl{complex} frequencies, while also providing a complete set of stable and passive basis vectors at \textsl{real} frequencies, and further connecting these physical attributes together in a consistent and origin-independent modal framework.
A basis of eigenmodes thereby provides both mathematical and physical insight into the underlying optical response of any finite scattering system.
As we will address in the next section, there is an added significance that the current model here contains loss and is thereby generally non-Hermitian, meaning eigenmodes need not be orthogonal.
However, the specific excitation of each eigenmode in the current model can still be found through the impact of reciprocity on the eigenmodes of any arbitrary system.
Moreover, Onsager's Reciprocity requires that the dyadic Green's function 
\index{Green's function}
and permittivity tensor be symmetric\cite{LandauLifshitzVol5}, albeit complex and not necessarily Hermitian.
\begin{align}
\bar{G}_{0}(\mathbf{r},\mathbf{r}')  = \bar{G}_{0}(\mathbf{r}',\mathbf{r}),  \quad\bar{G}_{0}= \bar{G}_{0}^{T},\quad {{\bar {\epsilon}}} = {{\bar {\epsilon}}}^{T}
\end{align}
Due to this symmetry, it is possible to write the overall operator of the eigenvalue equation (Equation~\ref{eq:eigenmode equation}) as a matrix in the normal form shown by Gantmacher~\cite{Gantmacher1959}.
This then shows that nondegenerate eigenmodes are orthogonal under unconjugated projections~\cite{Craven1969}.
\begin{align}
\int_{V}  {\mathbf{J}_v (\mathbf{r})} \cdot {\mathbf{J}_w (\mathbf{r})}   \; {\mathrm{dr}}^3= 0\,, \quad \mathrm{when}\;\,\lambda_v \neq \lambda_w \label{eq:this is responsible for orthogonality}
\end{align}
This pseudo-orthogonality of eigenmodes makes the excitation of any eigenmode determined through unconjugated dot products between the eigenmode and a driving field, analogous to the more familiar use of true projections when finding the excitation of orthogonal eigenmodes.
As such, despite eigenmodes being nonorthogonal, any given eigenmode's excitation amplitude $a_v$ is not dependent on the excitations of other eigenmodes: it is determined entirely by the given eigenmode's current distribution $\mathbf{J}_v$ and the driving field distribution $\mathbf{E_0}$.
\begin{align}
a_v = \frac{\int_{V}  {\mathbf{E}_v (\mathbf{r})} \cdot {\mathbf{J}_v (\mathbf{r})}   \; {\mathrm{dr}}^3}{\int_{V}  {\mathbf{J}_v (\mathbf{r})} \cdot {\mathbf{J}_v (\mathbf{r})}   \; {\mathrm{dr}}^3}
\end{align}
Turning now to our consideration of nanoparticle oligomers, this conclusion carries over to the eigenmodes of systems made from purely electric dipoles.
 \index{Oligomers}
An eigenmode $|v \rangle$ having electric dipoles $\mathbf{p}_v$ will satisfy Equation~\ref{eq:ED model} as:
\begin{align}
\mathbf{p}_v^{(i)} = \alpha_{\scriptscriptstyle E}^{(i)} \epsilon_0 \lambda_v \mathbf{p}_v^{(i)}  +  \alpha_{\scriptscriptstyle E}^{(i)} \sum \limits_{j\neq i} k^2 \bar{G}_{0}(\mathbf{r}_i,\mathbf{r}_j)\cdot \mathbf{p}_v^{(j)}
\label{eq:eig equation}
\end{align} 
This relationship between the dipole model 
\index{Dipole model}
and the current distribution model is not surprising given polarization and current are related as per Equation~\ref{eq:induced current}, where free currents are described in terms of polarization current.
Indeed, beyond accounting for the time derivative factor of $i \omega$ and the source dyadic $\bar{L}$, the dipole model is the same as the current model when the current distribution is described by a countable set of Dirac delta functions.

However, when introducing magnetic dipoles, we are describing the magnetic response of dielectric nanoparticles as magnetization.
Because of this, eigenmodes will be simultaneously constructed of both polarization and magnetization, where each has different units.
To address this difference of units, an initial approach was to separate the eigenmodes of electric and magnetic dipoles by introducing a term that describes the driving of electric dipoles by an applied magnetic field, and the driving of magnetic dipoles by an applied electric field~\cite{HopkinsPoddubny2013}.
This description will necessarily have cross terms accounting for interaction between the electric and magnetic eigenmodes, analogous to tensor bianisotropic polarizabilities.
However, while this is a full description of the dipole system, and it provides information on the resonances of electric and magnetic systems in the presence of each other, it {\sl does not} describe the simultaneous stable oscillations of both the electric {\sl and} magnetic dipoles, which are ultimately the resonances of the collective system.
To consider the eigenmodes and resonances of the  collective system, we must consider electric and magnetic dipole system simultaneously.
In this regard, it is first necessary to introduce relative scaling between the electric and magnetic dipoles to maintain fixed units of polarizability for the resulting eigenvalues: the magnetic dipoles can be scaled by a factor of ${c_0}^{-1}$, and the magnetic field by a factor of $\sqrt{\mu_0/\epsilon_0}$.
An eigenmode $|v \rangle$, having electric dipoles $\mathbf{p}_v$ and magnetic dipoles $\mathbf{m}_v$, will then satisfy the coupled electric and magnetic dipole model (Equation~\ref{eq:Dmodel}) as:
\begin{petit}
\begin{subequations}\label{eq:DmodelEig}
\begin{align}
 \lambda_v \, \mathbf{p}_v^{(i)} &=\,  (\bar{\alpha}_{{\scriptscriptstyle E}}^{(i)} \epsilon_{0} )^{-1} \mathbf{p}_v^{(i)}
-  \frac{k^{2}}{\epsilon_0}  \Big(\bar{G}_{0}(\mathbf{r}_{i},\mathbf{r}_{j})\cdot \mathbf{p}_v^{(j)}  +\nabla\times\bar{G}_{0}(\mathbf{r}_{i},\mathbf{r}_{j}) \cdot [c_0^{-1} \mathbf{m}_v^{(j)}]\Big)\\[2ex]
  \lambda_v \, [c_0^{-1} \mathbf{m}_v^{(i)}] &=\,
   (\bar \alpha_{{\scriptscriptstyle H}}^{(i)} \epsilon_0)^{-1}[c_0^{-1} \mathbf{m}_v^{(i)}]
-  \frac{k^{2}}{\epsilon_0}  \Big(\bar{G}_{0}(\mathbf{r}_{i},\mathbf{r}_{j}) \cdot [c_0^{-1} \mathbf{m}_v^{(j)}]
-\nabla\times\bar{G}_{0}(\mathbf{r}_{i},\mathbf{r}_{j}) \cdot\mathbf{p}_v^{(j)} \Big)
\end{align}%
\end{subequations}%
\end{petit}%
This expression describes a  matrix equation for eigenmodes of the electric and magnetic dipole system describing $N$ nanoparticles.
The associated $6N\times 6N$ matrix will not be symmetric when there is non-negligible coupling between the electric and magnetic dipoles, and therefore the corresponding eigenmodes will not maintain the pseudo-orthogonality analogous to that in Equation~\ref{eq:this is responsible for orthogonality} for currents.
Nonetheless, it worth acknowledging that this practice of replacing circulating current with magnetization remains analogous to the practice used to define permeability for conventional materials.
Therefore, the coupled electric and magnetic dipole model, including its resonances and their associated eigenmodes, should maintain the appropriate relationship to magnetic field to be the foundation for our analysis of all-dielectric nanoparticle oligomers.

\subsection{Modal Interference and Fano Resonances}
\label{sec:EPsDiscussedBriefly}
\begin{figure}
\centerline{\includegraphics[width=0.95\textwidth]{{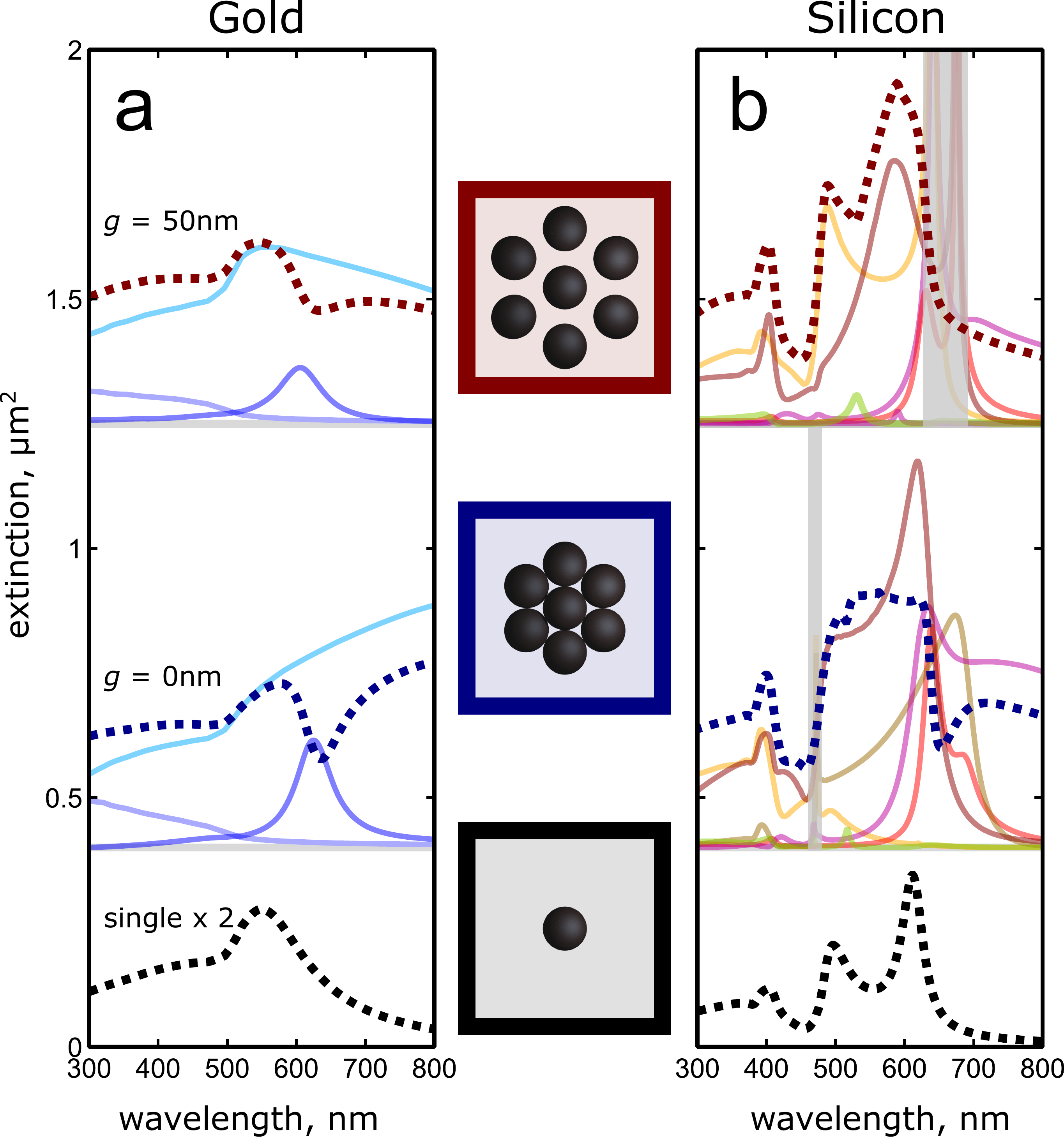}}}
\caption{(Dashed lines) Extinction spectra of (a) gold and (b) silicon heptamers, simulated using the dipole model Equation~\ref{eq:Dmodel} and showing the role of eigenmode interference in producing Fano resonances.
(Solid lines) Overlaid direct terms to extinction, as per Equation~\ref{eq:decomp}, for all three excited eigenmodes of the gold heptamers, and the six most dominant eigenmodes for the silicon heptamers.
The gray regions correspond to wavelength bands near exceptional points.  Both gold and silicon nanospheres are 150~nm in diameter.}
\label{fig:heptamers}
\end{figure}

We begin by noting that both the current model and the dipole model describe open, radiative systems.
As such, even in the absence of material loss, the system does still exhibit radiation losses and is generally, therefore, {\it non-Hermitian}.
The immediate consequence of this non-Hermicity is that the eigenmodes we have defined in Equations~\ref{eq:eigenmode equation}, \ref{eq:eig equation} and \ref{eq:DmodelEig} are not necessarily orthogonal.
To recognize the effect of this nonorthogonality, and its relation to the Fano resonances, 
\index{Fano resonance}
we can refer to the extinction cross-section, which describes the total loss in a system, and can be constructed the sum of the scattering and absorption cross-sections.
The scattering cross-section can be calculated from the integral of the far-field scattered power, such as derived by Merchiers {\it et al.}\cite{Merchiers2007} for the case of electric and magnetic dipoles.
\begin{align}
\sigma_\mathrm{s} =&  \frac{k}{{\epsilon_0}^2 I_0} \mathrm{Im}\bigg[ \sum \limits_i  \epsilon_0\mathbf{E_0}^*(\mathbf{r}_i) \cdot  \mathbf{p}_i + {\epsilon_0}{\mu_0}\mathbf{H_0}^*(\mathbf{r}_i) \cdot  \mathbf{m}_i \nonumber \\
&+\mathbf{p}_i^*\cdot\Big (\frac{ i k^3}{6 \pi}  + (\bar\alpha_{\scriptscriptstyle E}^{(i)})^{-1}\Big) \mathbf{p}_i +{\epsilon_0}{\mu_0} \mathbf{m}_i^* \cdot \Big(\frac{ i k^3}{6 \pi}   + \big(\bar\alpha_{\scriptscriptstyle H}^{(i)})^{-1}\Big)  \mathbf{m}_i \bigg]
\end{align}
where $I_0$ is associated with the average intensity of the excitation such that it relates the area of a cross-section $\sigma$ to the total power $P =\frac{1}{2}\sqrt{\frac{\mu_0}{\epsilon_0}}\, I_0 \, \sigma$.
The absorption cross-section can be calculated from the closed surface integral of the total Poynting vector field around the scattering object, or by calculating the Ohmic losses of the internal electric and magnetic field\cite{Kern2010}.
\begin{align}
\sigma_\mathrm{a} =& \frac{-k}{{\epsilon_0}^2 I_0} \mathrm{Im}\bigg[ \mathbf{p}_i^*\!\cdot\!\Big (\frac{ i k^3}{6 \pi}  + (\bar\alpha_{\scriptscriptstyle E}^{(i)})^{-1}\Big) \mathbf{p}_i +{\epsilon_0}{\mu_0} \mathbf{m}_i^* \!\cdot \! \Big(\frac{ i k^3}{6 \pi}  + \big(\bar\alpha_{\scriptscriptstyle H}^{(i)})^{-1}\Big)  \mathbf{m}_i \bigg]
\end{align}
The extinction cross-section, as the sum of absorption and scattering, can then be written as:
\begin{align}
\sigma_\mathrm{e} =& \frac{k}{\epsilon_0 I_0} \mathrm{Im}\bigg[ \sum \limits_i  \mathbf{E_0}^*(\mathbf{r}_i) \cdot  \mathbf{p}_i + \mu_0\mathbf{H_0}^*(\mathbf{r}_i) \cdot  \mathbf{m}_i \bigg]
\end{align}
The equivalent expression for extinction in the current model, Equation~\ref{eq:current equation}, can be found\cite{HopkinsPoddubny2016} by expressing a current distribution as a discretized distribution of point dipoles that each occupy an infinitesimal  volume $\mathrm{dr}^3$.
\begin{align}
\sigma_{\mathrm{e}}  &= \frac{1}{I_0}\sqrt{\frac{\mu_0}{\epsilon_0}} \; \mathrm{Re}\Big [ \int_{V_s} \mathbf{E}_0^* \cdot \mathbf{J}  \; \mathrm{dr}^3 \Big ] \label{eq:extinction definition}
\end{align}
Notably, this expression corresponds to the volume integral of the time-averaged power imparted locally by any field distribution $\mathbf{E_0}$ on the currents $\mathbf J$, though written here as a cross-section.
So let us now acknowledge that an arbitrary applied field and its induced currents can be defined in terms of a linear superposition of the eigenmodes: 
\begin{align}
 \mathbf{E}_{0} &= \sum \limits _v a_v \lambda_v {\mathbf{J}_v}  \quad \Rightarrow \quad \mathbf{J} = \sum \limits _v a_v {\mathbf{J}_v} \label{eq:first J}
\end{align}
We are then able to rewrite the extinction (Equation~\ref{eq:extinction definition}) in terms of eigenmodes and eigenvalues.
Moreover, we can divide the extinction into two contributions: {\it direct terms} that provide contributions to extinction from individual eigenmodes, and also {\it interference terms} coming from the overlap between different eigenmodes.
\index{Eigenmode decomposition}
\begin{petit}
\begin{align}
\sigma_{\mathrm{e}}  &=  \frac{1}{I_0} \sqrt{\frac{\mu_0}{\epsilon_0}}\sum \limits_{v} \Bigg ( \; \underset{\mathrm{direct\;terms}}{\underbrace{\mathrm{Re}[\lambda_v]  \int_{V} |a_v|^{2}|\mathbf{J}_v|^{2}  \; \mathrm{dr}^3 }}\; + \sum \limits_{w \neq v}\;\underset{\mathrm{interference\;terms}}{\underbrace{\mathrm{Re}\Big[ a_v^{*} a_w \lambda_v^*  \int_{V} \mathbf{J}_v^{*}\cdot \mathbf{J}_w \; \mathrm{dr}^3 \Big]}}\Bigg) 
\label{eq:decomp}
\end{align}
\end{petit}
\noindent The direct terms are always greater than zero due to the inscribed passivity of eigenmodes, and interference terms would notably all be zero for orthogonal eigenmodes.
As we will now explain, the existence of nonzero interference terms, and thereby the presence of nonorthogonal eigenmodes, is required for Fano resonances to exist.
Moreover, a given eigenmode's excitation amplitude, being the $a_v$ coefficients of Equation~(\ref{eq:first J}), is determined  independent from the excitations of other eigenmodes: \mbox{$a_v \!=\! \frac{\int  {\mathbf{E_0}} \cdot {\mathbf{J}_v}    \, {\mathrm{dr}}^3}{\int{\mathbf{J}_v} \cdot {\mathbf{J}_v}    \, {\mathrm{dr}}^3}$}.
This conclusion follows directly from the pseudo-orthogonality in Equation~\ref{eq:this is responsible for orthogonality}.
The only way an interaction between two or more eigenmodes affects the extinction cross-section is, therefore, through interference terms, meaning that Fano resonances must be described by the nonorthogonality of eigenmodes.
This is indeed the conclusion of earlier works on the Fano resonances arising in nanoparticle oligomers\cite{HopkinsPoddubny2013, Forestiere2013}, 
 \index{Oligomers}
 and is in good accordance with Fano resonances that  have been proposed and observed to occur between modes that are directly driven by the incident field\cite{Frimmer2012, Lovera2013}.
As an example, we consider heptamers made from gold nanospheres shown in Figure~\ref{fig:heptamers}a.
By performing an eigenmode decomposition of this system, we are able to show the direct terms of extinction coming from the dominant eigenmodes, in addition to the total extinction.
This means the difference between total extinction and the sum of direct terms are the interference terms coming from eigenmode overlap.
For the case of the gold heptamer, we have a typical scenario of the Fano resonances: the overlap of a broad resonance and a sharp resonance, and coupling that leads to destructive interference.
This gold heptamer thereby shows a classical example of a Fano resonance\cite{Miroshnichenko2010}, where the interaction between resonances is coming from eigenmode nonorthogonality.

However, as seen in Figure~\ref{fig:heptamers}b, the situation becomes dramatically more complicated for a silicon heptamer.
The number of eigenmodes of this system is much higher, because of both the additional magnetic dipoles and because electric-magnetic dipole coupling ({\it cf.} Equation~\ref{eq:EMcoupling}) allows each dipole to be polarized parallel to the propagation direction\cite{HopkinsFilonovPRB2015}.
The first consequence of this increased eigenmode count is that there are many more signatures of interference phenomena occurring across the extinction spectra.
However, the second consequence is that we have to accept the formation of  {\it exceptional points}.
An exceptional point refers to a point degeneracy of two or more eigenvalues when their eigenmodes simultaneously become linearly dependent, subsequently corresponding to a reduction in the dimension of the span of eigenmodes\cite{Heiss2001, Dembowski2001, Heiss2012}.
Exceptional points can exist even in simple plasmonic and dielectric oligomer systems at complex frequencies\cite{HopkinsFilonovPRB2015}, but can generally be expected to occur more regularly when there are more interacting eigenmodes.
For the discussion here, we need to recognize that the excitation magnitudes of coalescing eigenmodes can diverge in the vicinity of an exceptional point, given a component of the driving field is gradually becoming orthogonal to the eigenmodes that span it.  
However, a full discussion of the properties of exceptional points will not be covered, and we have therefore attempted to divert attention away from the direct extinction terms of individual eigenmodes that diverge as they near exceptional points in Figure~\ref{fig:heptamers}b.
The reason we observe such features is likely because the number of eigenmodes that are excited by a normally-incident plane wave in a dielectric oligomer can be more than double that for a plasmonic oligomer.
Assuming a normally-incident plane wave, the number of excitable eigenmodes  in these rotationally-symmetric plasmonic oligomers is relatively contained: two degenerate pairs of eigenmodes for each ring of nanoparticles  and one further pair if there is central nanoparticle\cite{HopkinsPoddubny2013}.
On the other hand, the coupling between electric and magnetic dipoles in dielectric oligomers removes such eigenmode restrictions, making the number of excitable eigenmodes increase with the number of nanoparticles in each ring\cite{HopkinsFilonovPRB2015}.
This means that even simple arrangements of dielectric nanoparticles can be used to achieve the behavior of more complicated plasmonic oligomers, an example of which will be investigated in the next section.

Finally, as an addendum, we should recognize that there is a degree of freedom in how to attribute extinction to each given eigenmode.
In particular, rather than separating direct and interference terms as per Equation~\ref{eq:decomp}, Frimmer {\it et al.}\cite{Frimmer2012} attributed the extinction associated with a given eigenmode to the projection of this eigenmode onto the complete incident field.
\index{Eigenmode decomposition}
\begin{align}
\sigma_{\mathrm{e}}  &=  \frac{1}{I_0} \sqrt{\frac{\mu_0}{\epsilon_0}}\sum \limits_{v} \;{{\mathrm{Re}\Big[ a_v \big( \int_{V} \mathbf{E_0}^{*}\cdot \mathbf{J}_v \; \mathrm{dr}^3\big) \Big]}} 
\label{eq:altdecomp}
\end{align}
In this approach, Fano resonances appear as negative extinction from an eigenmode\cite{Frimmer2012}, as per negative interference terms overpowering the positive direct terms.
This decomposition ultimately contains the same interactions and nonorthogonality of eigenmodes, however, it is still important to emphasize the role of modal overlap in producing Fano resonances: the eigenmodes themselves are the basis vectors that produce only positive extinction, and negative extinction only arises from the overlap between eigenmodes.
The choice of deconstructing extinction using Equation~\ref{eq:decomp} or \ref{eq:altdecomp}, is effectively a question of whether or not to separate interference terms.

\subsection{Eigenmodes of Nanoparticle Dimers}
\index{Dimer}
Up until now, we have calculated eigenmodes and eigenvalues through direct numerics, which does not necessarily offer much insight on how collective resonances have formed.
Here we will instead present  a method for calculating eigenmodes by subdividing the optical response of a complete structure into a set of constituent subsystems, and then adding the appropriate coupling between these subsystems.
This approach was utilized for a nanoparticle trimer\cite{HopkinsFilonovPRB2015} when employing the dipole model, and it was possible to derive even {\sl analytical} expressions for eigenmodes.
Here we will instead present derivations for the resonances of two closely-spaced nanoparticles, so-called {\it dimers}, owing to recent interest in silicon nanoparticle dimers that exhibit behavior such as Fano resonances, directionality and magnetic near-field enhancement~\cite{Yan2015, Bakker2015, Zywietz2015}.
There has also been an interest in hybrid dimers consisting both high-index dielectric nanoparticles {\sl and} plasmonic nanoparticles, including predictions of antiferromagnetic behavior~\cite{Miroshnichenko2011}.
In this regard, symmetric dimers have previously been addressed~\cite{Merchiers2007}, but {\sl asymmetry} is necessary to describe some of the above mentioned works.
Additionally, an asymmetric dimer also resembles the more common situation of a particle on a substrate, where the substrate can be treated as a virtual image of the particle.

\begin{figure*}[!t]
\centerline{\includegraphics[width=\textwidth]{{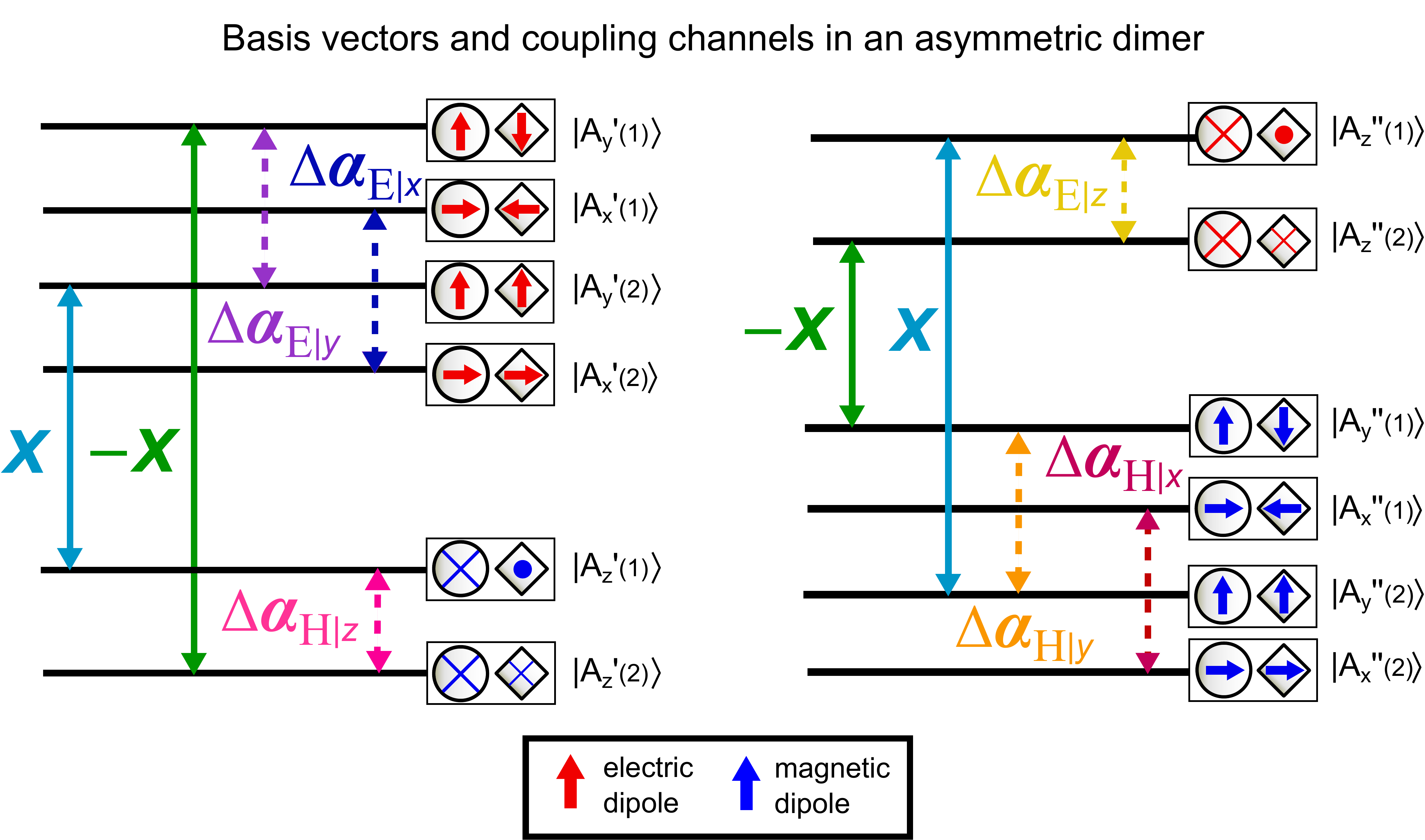}}}
\caption{Diagram showing the complete set of basis vectors, and the coupling channels between them, for the optical response of an asymmetric dimer.
Individual nanoparticles are considered to exhibit both electric (red) and magnetic (blue) dipole moments.  }
\label{fig:asymmetricFull}
\end{figure*}

In the coming derivations we will simplify our analysis by considering only diagonal tensor dipole polarizabilities with three primary axes ($x$, $y$ and $z$ axes), and neglect both anisotropic and bianisotropic terms.
\begin{align}
\bar{\alpha}_{{\scriptscriptstyle E}}^{(i)} =
\left(\begin{array}{ccc}
\alpha_{{\scriptscriptstyle E}|x}^{(i)}&&\\
& \alpha_{{\scriptscriptstyle E}|y}^{(i)}& \\
& & \alpha_{{\scriptscriptstyle E}|z}^{(i)}
\end{array}\right)\label{eq:tensorPol}
\end{align}
and similarly for $\bar{\alpha}_{{\scriptscriptstyle H}}$.
We then use matrix notation to denote eigenmodes as vector concatenations of the associated dipole moments:
\begin{align}
 | v \rangle = \left ( \begin{array}{c} \mathbf{p}_v^{\scriptscriptstyle(1)} \\[0.5ex] \mathbf{p}_v^{\scriptscriptstyle(2)} \\[0.5ex] c_0^{-1} \mathbf{m}_v^{\scriptscriptstyle(1)} \\[0.5ex] c_0^{-1}  \mathbf{m}_v^{\scriptscriptstyle(2)}\end{array} \right),\quad
 \lambda | v \rangle = \left ( \begin{array}{c} \mathbf{E_0}_v(\mathbf{r}_1) \\[0.5ex] \mathbf{E_0}_v(\mathbf{r}_2)\\[0.5ex]\sqrt{\frac{\mu_0}{\epsilon_0}} \mathbf{H_0}_v(\mathbf{r}_1) \\[0.5ex] \sqrt{\frac{\mu_0}{\epsilon_0}} \mathbf{H_0}_v(\mathbf{r}_2)\end{array} \right)
 \label{eq:scaling}
\end{align}
The relative scaling of the electric and magnetic dipole moments, and of the electric and magnetic applied fields, is again used to give the eigenvalues units corresponding to an inverse polarizability.
Using the state notation in Equation~\ref{eq:scaling},  the eigenmode equation (Equation~\ref{eq:DmodelEig}) can be compactly written in terms of both a polarizability operator $\boldsymbol{\hat{\alpha}}$ and a Green's function operator $\boldsymbol{ \hat{\mathcal{G}}}$.
\begin{align}
\lambda\,| v \rangle   = \Big(\boldsymbol{\hat{\alpha}}^{-1} - \boldsymbol{\hat{\mathcal{G}}} \Big)  | v \rangle
\label{eq:matrixAsym}
\end{align}
\index{Dipole model}
This allows us to find simplified matrix versions of the $\boldsymbol{\hat{\alpha}}$ and $\boldsymbol{\hat{\mathcal{G}}}$ operators for subsets of eigenmodes that transform according to a single irreducible representation of the dimer's symmetry point group.
In particular, when considering the general case of an asymmetric dimer with achiral nanoparticles, we reduce the set of geometric symmetry operations to those of a  $C_{1h}$  point group.
We can then define basis vectors of the dimer's optical response to be the eigenmodes of the $ \boldsymbol{\hat{\mathcal{G}}}$ operator in the absence of electric to magnetic dipole coupling.
These basis vectors are those shown in Fig.~\ref{fig:asymmetricFull}, which are named according to their irreducible representations in the ${C_{1h}}$ point group, but with an additional subscript $x,y,z$ to distinguish the orientation of each such basis vector and account for anisotropic nanoparticles as per Equation~\ref{eq:tensorPol}.
Notably these basis vectors are more commonly associated with the eigenmodes of a symmetric dimer, but they are also practical for an asymmetric dimer because they still form a complete basis for its optical response.
Moreover, because they remain eigenmodes of the Green's function operator in the absence of electric to magnetic coupling, we only need to consider coupling channels between basis vectors created by the $\boldsymbol{\hat{\alpha}}^{-1}$ operator, and channels due to electric to magnetic coupling.
In  Fig.~\ref{fig:asymmetricFull}, we use the convention where a dashed line represents a coupling channel from the dissimilarity of nanoparticles in the $\boldsymbol{\hat{\alpha}}^{-1}$ polarizability operator and an unbroken line represents the bianisotropic coupling channels in the $ \boldsymbol{\hat{\mathcal{G}}}$ Green's function operator.
The magnitudes of these coupling channels can be determined from the polarizabilities  of each nanoparticle, and from the dipole model in Equation~\ref{eq:Dmodel}.
\begin{align}
\Delta\alpha^{-1} & =\frac{1}{2}\Big(\frac{1}{\alpha^{(1)}}  -\frac{1}{\alpha^{(2)}}\Big) \\
 X_{\uparrow} = -X_{\downarrow} \equiv X &= \frac{e^{i k R }}{4 \pi \epsilon_0 R} \Big( k^{2} + \frac{ik}{R}\Big)
\end{align}
We can also define the eigenvalues $\gamma$ of the Green's function operator independently of the polarizability operator.
\begin{align}
& \gamma_{ 1} =  \frac{e^{i k R }}{4 \pi \epsilon_0 R} \Big( k^{2} + \frac{ik}{R}- \frac{1}{R^2}\Big) \\
&  \gamma_{2} =  -\frac{ e^{i k R }}{2 \pi \epsilon_0  R} \Big( \frac{ik}{R}- \frac{1}{R^2}\Big)
\end{align}
where $R = |\mathbf{r}_1-\mathbf{r}_2|$ is the distance between dipoles.
Using this approach, the operator $\big(\boldsymbol{\hat{\alpha}}^{-1} - \boldsymbol{\hat{\mathcal{G}}} \big)$  can be written as a $2 \times 2$ matrix for each of the $\mathrm{A'_x}$ and $\mathrm{A''_x}$ basis vectors:
\begin{align}
\boldsymbol{\hat{\alpha}}^{-1} -  \boldsymbol{\hat{\mathcal{G}}} &\;=\; {\displaystyle \frac{ 1}{ \epsilon_0}}\left(
 \begin{array}{cc}
{ \langle \alpha^{-1}\rangle } + \epsilon_0 \gamma
\;&\; { \Delta\alpha^{-1}} \\[1ex]
{ \Delta\alpha^{-1}}
 &{ \langle \alpha^{-1}\rangle -\epsilon_0  \gamma }
 \end{array}\right)
 \label{eq:simplified}
\end{align}
where:
\begin{align*}
\Delta\alpha^{-1} & =\frac{1}{2}\Big(\frac{1}{\alpha^{(1)}}  -\frac{1}{\alpha^{(2)}}\Big)
\\
\langle \alpha^{-1}  \rangle & =\frac{1}{2}\Big(\frac{1}{\alpha^{(1)}}  +\frac{1}{\alpha^{(2)}}\Big)
\end{align*}
The particular polarizability and component that $\alpha$ refers to has been left intentionally vague so we can reuse the notation; the particular values we use are listed in Tables~\ref{tab:simplified} and \ref{tab:quartic}.
\begin{table}[t]
\centering
 \begin{tabular}{c || c | c | c | c  }
&\;\, \large$|\Gamma^{(a)} \rangle \;\,$   &\;\, \large $ |\Gamma^{(b)} \rangle\;\, $   & \large$ \;\;  \alpha^{(1)}  \;\,$  &\large $ \;\; \alpha^{(2)} \;\, $  \\[0.75ex]\hline \hline
& &   & &   \\ [-1.75ex]
\large$\mathrm{A'_x} \;\,$  & $|\mathrm{A'_{x\scriptscriptstyle(1)}}\rangle$ & $|\mathrm{A'_{x(2)}}\rangle$ &    $\alpha_{{\scriptscriptstyle E}|x}^{(1)}$  &    $\alpha_{{\scriptscriptstyle E}|x}^{(2)}$  \\[1.5ex]
\large$\mathrm{A''_x} \;\,$ &$|\mathrm{A''_{x\scriptscriptstyle(1)}}\rangle$ &$|\mathrm{A''_{x\scriptscriptstyle(2)}}\rangle$ &   $\alpha_{{\scriptscriptstyle H}|x}^{(1)}$  &    $\alpha_{{\scriptscriptstyle H}|x}^{(2)}$  \\[1.5ex]
\end{tabular}
\caption{The combinations of basis vectors and polarizabilities that define pairs of eigenmodes for the $\mathrm{A'_x}$ and $\mathrm{A''_x}$ responses, as described by Equation~\ref{eq:AX}}.
\label{tab:simplified}
\end{table}
\\
For the remaining basis vectors of the complete dimer, the bianisotropic coupling channels require a $4\times4$ matrix in order to describe the $\boldsymbol{\hat{\alpha}}^{-1} -  \boldsymbol{\hat{\mathcal{G}}} $ operator.
\begin{align}
\boldsymbol{\hat{\alpha}}^{-1} -  \boldsymbol{\hat{\mathcal{G}}} &\;=\;{  {\displaystyle \frac{ 1}{ \epsilon_0}}} \left( \begin{array}{cccc}
 \langle \alpha_{{\scriptscriptstyle E}}^{-1}\rangle  + \epsilon_0\gamma_{ 1}
\;&\; { \Delta\alpha_{{\scriptscriptstyle E}}^{-1}}
&
&\epsilon_0 X \\[1ex]
{ \Delta\alpha_{{\scriptscriptstyle E}}^{-1}}
 &{ \langle \alpha_{{\scriptscriptstyle E}}^{-1}\rangle - \epsilon_0\gamma_{ 1} }
 &-\epsilon_0 X
&\\ [1ex]
\;&\; \epsilon_0 X
& { \langle \alpha_{{\scriptscriptstyle H}}^{-1}\rangle +\epsilon_0\gamma_{ 1} }
\;&\; { \Delta\alpha_{{\scriptscriptstyle H}}^{-1}} \\[1ex]
-\epsilon_0 X
&
&{ \Delta\alpha_{{\scriptscriptstyle H}}^{-1}}
 &{ \langle \alpha_{{\scriptscriptstyle H}}^{-1}\rangle - \epsilon_0\gamma_{ 1}}
\end{array}\right)
\label{eq:fullMatrix}
\end{align}
Here the indices of the polarizabilities are different for the $\mathrm{A'}$ and $\mathrm{A''}$ basis vectors as prescribed in Table~\ref{tab:quartic}.
\begin{table}[b]
\centering
 \begin{tabular}{c || c | c }
& \large $\;\,   A' \;\,$   &\large $\;\, A'' \;\,$ \\[0.75ex]\hline \hline
& &   \\ [-1.75ex]
\large$\alpha_E\;$    & \large $\;\alpha_{{\scriptscriptstyle E}|y} \;$ & \large$\;\alpha_{{\scriptscriptstyle E}|z}\;$\\[1.5ex]
\large$\alpha_H\;$    & \large$\;\alpha_{{\scriptscriptstyle H}|z}\;$ & \large$\;\alpha_{{\scriptscriptstyle H}|y}\;$ \\[1.5ex]
\large$X\;$    & \large$\;X\;$ & \large$\;-X\;$ \\[1ex]
\end{tabular}
\caption{The combinations of polarizability axes and coupling channel for {Equations~\ref{eq:begin}--\ref{eq:end}} that provide the eigenvalues and eigenmodes of the $\mathrm{A'}$ and $\mathrm{A''}$ irreducible representations according to Equation~\ref{eq:eigenvaluesFull} and Equations~\ref{eq:eigmodeBegin}--\ref{eq:eigmodeEnd}}
\label{tab:quartic}
\end{table}
Given we now have equivalent matrix expressions for the independent subspaces of the  $\boldsymbol{\hat{\alpha}}^{-1} -  \boldsymbol{\hat{\mathcal{G}}} $ operator, shown in Equations~\ref{eq:simplified} and \ref{eq:fullMatrix}, we can find the dimer's eigenmodes and eigenvalues such that they satisfy:
\begin{align}
\lambda_{v}^{(i)}| v_i \rangle  = (\boldsymbol{\hat{\alpha}}^{-1} - \boldsymbol{\hat{\mathcal{G}}})| v_i \rangle
\end{align}
In other words, the complete set of eigenmodes can be found from the eigendecompositions of $2\times2$ and $4\times4$ matrices.
Moreover, for the case of the $2\times2$ matrices for $\mathrm{A'_x}$ and $\mathrm{A''_x}$ response space, the eigenmodes and eigenvalues are of the form:
\begin{subequations} \label{eq:AX}
\begin{align}
 |v_{\pm} \rangle & =  \Big( \epsilon_0\gamma_2   \pm \delta_v \Big ) |\Gamma^{(b)} \rangle
-\Delta\alpha^{-1}  |\Gamma^{(a)} \rangle   \\
\lambda_{\pm} &= \frac{\langle \alpha^{-1}  \rangle \mp \delta_v}{\epsilon_0}
\end{align}
\end{subequations}
where the specific values to use for each polarizability $\alpha$ and the basis vectors $|\Gamma \rangle  $ are written in Table~\ref{tab:simplified}, and we have also defined a function $\delta$ to explicitly quantify the difference between eigenmodes:
\begin{align}
\delta_v & = \frac{1}{\alpha^{(1)}\alpha^{(2)}}
\sqrt{\Big(  \frac{\alpha^{(1)} - \alpha^{(2)}}{2} \Big)^{2} - \Big( \alpha^{(1)}\alpha^{(2)} \epsilon_0 \gamma  \Big) ^{2}}
\end{align}
Equation~\ref{eq:AX} gives general expressions for the eigenmodes and eigenvalues of the $\mathrm{A'_x}$ and $\mathrm{A''_x}$ response spaces.
Notably, as per the discussion in Section~\ref{sec:EPsDiscussedBriefly}, there is an {\it exceptional point}\cite{Heiss2012} occurring when $\delta = 0$; a degeneracy of eigenvalues combined with a linear dependency of the eigenmodes.

\begin{figure}
\centerline{\includegraphics[width=\textwidth]{{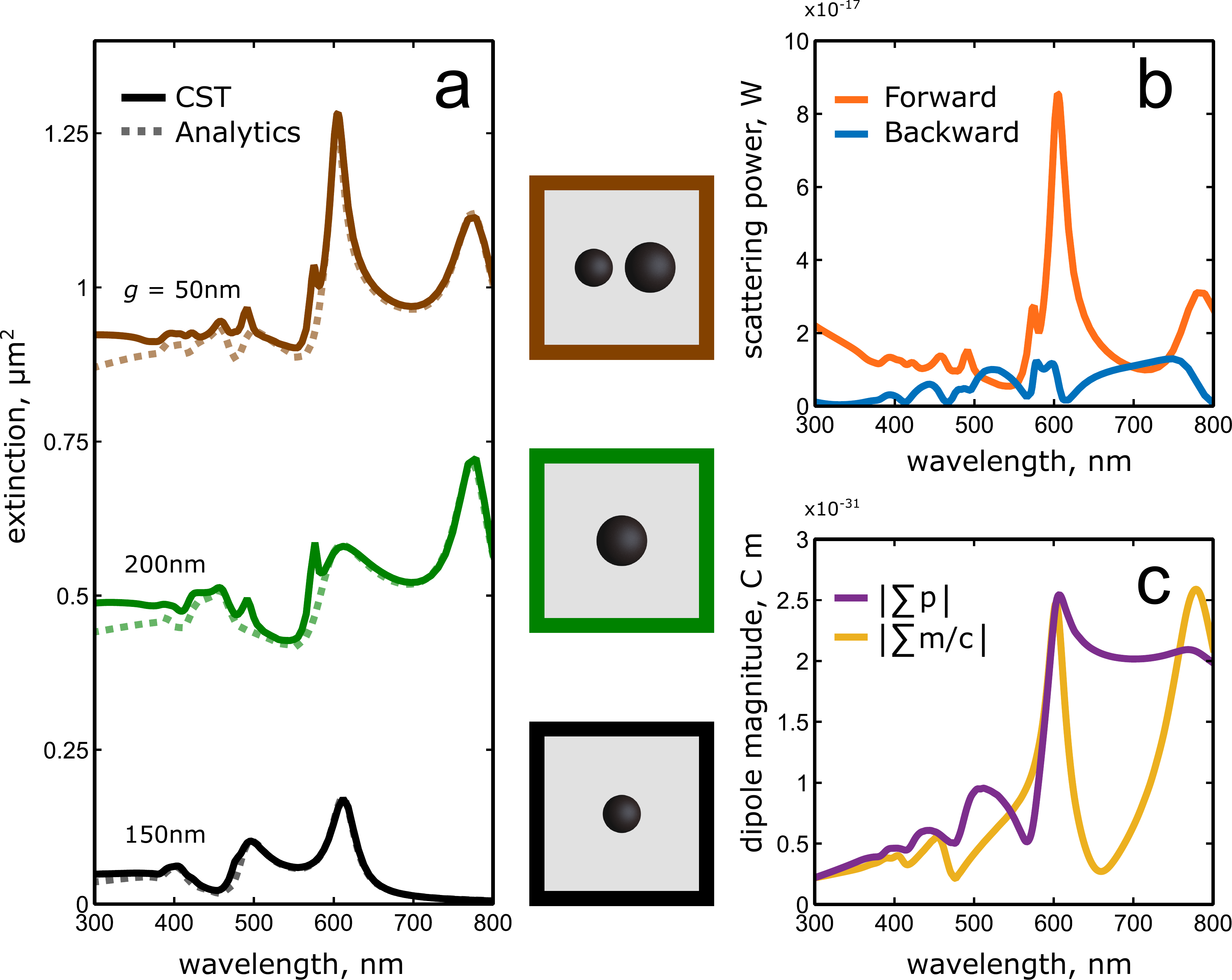}}}
\caption{(a) Calculated extinction spectra for a 150~nm and a 200~nm diameter silicon nanosphere, and the combined dimer when separated by $g=50~\mathrm{nm}$, showing good agreement between CST and the analytical result for eigenmodes and eigenvalues derived in this section.  (b) The forward and backward scattered power in $\pi/6$ solid angles of a plane wave with unit amplitude (1~V/m), and (c) the associated equality of electric and magnetic dipoles in the dimer.}
\label{fig:dimer1}
\end{figure}

To find the remaining eigenvalues of the dimer, belonging to the $(\boldsymbol{\hat{\alpha}}^{-1} -  \boldsymbol{\hat{\mathcal{G}}})$ operator in Equation~\ref{eq:fullMatrix}, we solve for the roots of the associated characteristic equation.
The general form of the characteristic equation for both $\mathrm{A'}$ and $\mathrm{A''}$ response spaces will be:
\begin{align}
\lambda^4 + B \lambda^3 + C \lambda^2 + D \lambda+ E = 0 
\label{eq:quartic}
\end{align}
where each coefficient is given by Equations~\ref{eq:begin}--\ref{eq:end}.
\begin{petit}
\begin{align}
B & =  - {\epsilon_0^{-1}}\Big( \langle \alpha_{\scriptscriptstyle E}^{-1}  \rangle +  \langle \alpha_{\scriptscriptstyle H}^{-1}  \rangle\Big)  \label{eq:begin} \\
C & = \epsilon_0^{-2}\Big( {\langle \alpha_{\scriptscriptstyle E}^{-1}  \rangle}  \langle \alpha_{\scriptscriptstyle H}^{-1}\rangle
+ \frac{1}{\alpha_{\scriptscriptstyle E}^{(1)}\alpha_{\scriptscriptstyle E}^{(2)} }
+ \frac{1}{\alpha_{\scriptscriptstyle H}^{(1)}\alpha_{\scriptscriptstyle H}^{(2)}}
 +2 \epsilon_0^2 \big ({X^2} -\gamma_1^2\big)
 \Big)\\
D & =\epsilon_0^{-3} \Big( {\langle \alpha_{\scriptscriptstyle E}^{-1}  \rangle} \Big[ \epsilon_0^{2} \gamma_1^2 - \frac{1}{\alpha_{\scriptscriptstyle H}^{(1)}\alpha_{\scriptscriptstyle H}^{(2)}} \Big]
+  {\langle \alpha_{\scriptscriptstyle E}^{-1}  \rangle} \Big[ \epsilon_0^{2} \gamma_1^2 - \frac{1}{\alpha_{\scriptscriptstyle E}^{(1)}\alpha_{\scriptscriptstyle E}^{(2)}} \Big ]
- \epsilon_0^2{X^2} \Big [ {\langle \alpha_{\scriptscriptstyle E}^{-1}  \rangle} +  \langle \alpha_{\scriptscriptstyle H}^{-1}  \rangle \Big]
 \Big)  \\
E & =\epsilon_0^{-4} \Big(
 \frac{1}{\alpha_{\scriptscriptstyle E}^{(1)}\alpha_{\scriptscriptstyle E}^{(2)} \alpha_{\scriptscriptstyle H}^{(1)}\alpha_{\scriptscriptstyle H}^{(2)}}
 +  \epsilon_0^2{X^2} \Big [\frac{1}{\alpha_{\scriptscriptstyle E}^{(1)}\alpha_{\scriptscriptstyle H}^{(2)} } + \frac{1}{\alpha_{\scriptscriptstyle H}^{(1)}\alpha_{\scriptscriptstyle E}^{(2)}} \Big]\nonumber\\ &\qquad\qquad\quad
-  \epsilon_0^2{\gamma_1^2} \Big [\frac{1}{\alpha_{\scriptscriptstyle E}^{(1)}\alpha_{\scriptscriptstyle E}^{(2)} } + \frac{1}{\alpha_{\scriptscriptstyle H}^{(1)}\alpha_{\scriptscriptstyle H}^{(2)}} \Big]
+ \epsilon_0^{4}{ (X^2 - \gamma_1^2)}^2
\Big)
\label{eq:end}
\end{align}
\end{petit}
The roots of quartic equations like Equation~\ref{eq:quartic} are known for general coefficients.
Indeed, we write the resulting eigenvalues using four different combinations of plus and minus signs, $\pm_{_{(1)}}$ and $\;\pm_{_{(2)}}$:
\begin{petit}
\begin{subequations}
\begin{align}
\lambda &\;=\; -\frac{1}{4}B
\; \pm_{_{(1)}}\; \frac{1}{2}\sqrt{\frac{1}{4}B^2-C+h\,}
\;\;\pm_{_{(2)}}\;\frac{1}{2}\sqrt{\frac{1}{2}B^2-C-h
\;\; \pm_{_{(1)}}\;   \frac{-\frac{1}{4}B^3 + B C - 2 D}{\sqrt{\frac{1}{4}B^2-C+h\,}}} \\[2ex]
 \lambda &\;\rightarrow\; -\frac{1}{4}B
\;\;\pm_{_{(2)}}\;\frac{1}{2}\sqrt{\frac{3 }{4}B^{2} -2C \; \pm_{_{(1)}}\;  2 \sqrt{h^2 - 4 E}}\,,\quad \mathrm{when}\quad\frac{1}{4}B^2-C+h = 0
\\[1ex]
\nonumber & \mathrm{where:}\\
&\nonumber\quad  h = \frac{1}{3}\Bigg(C +  q\; \bigg[\frac{p + \sqrt{p^2-4 q^3}}{2}\bigg]^{-1/3}+  \bigg[ \frac{p + \sqrt{p^2-4 q^3}}{2}\bigg]^{1/3} \Bigg)\\
&\nonumber\quad p= 2 C^3 - 9 B C D + 27  D^2 + 27 B^2 E - 72 C E\\
&\nonumber\quad q  = C^2 - 3 B D + 12 E
\end{align}\label{eq:eigenvaluesFull}
\end{subequations}
\end{petit}
Using these eigenvalues we can now find the dipole moment profiles of the eigenmodes themselves.
We begin by writing an eigenmode as a linear combination of the associated $\mathrm{A'}$ and $\mathrm{A''}$ basis vectors,
\begin{align}
| v_i \rangle = \Bigg \{
\begin{array}{cc}
 a_i |\mathrm{A'_{\scriptscriptstyle(y2)}}\rangle + b_i |\mathrm{A'_{\scriptscriptstyle(y1)}}\rangle
 + c_i |\mathrm{A'_{\scriptscriptstyle(z2)}}\rangle + d_i |\mathrm{A'_{\scriptscriptstyle(z1)}}\rangle & \quad (A')\\[1ex]
 a_i |\mathrm{A''_{\scriptscriptstyle(y2)}}\rangle + b_i |\mathrm{A''_{\scriptscriptstyle(y1)}}\rangle
 + c_i |\mathrm{A''_{\scriptscriptstyle(z2)}}\rangle + d_i |\mathrm{A''_{\scriptscriptstyle(z1)}}\rangle & \quad (A'')
 \end{array}
 \label{eq:linearComb}
\end{align}
We can then write down the relationships between the coefficients of these $\mathrm{A'}$ and $\mathrm{A''}$ basis vectors using the coupling channels depicted by Figure~\ref{fig:asymmetricFull} and expressed in Equation~\ref{eq:fullMatrix}.
\begin{subequations}
\begin{align}
X d_i  &= \Big(\lambda_{i}-{\textstyle \frac{1} {\epsilon_0}}\langle \alpha_{{\scriptscriptstyle E}}^{-1}\rangle  + \gamma_1\Big)a_i
- {\textstyle \frac{1} {\epsilon_0}}  \Delta\alpha_{{\scriptscriptstyle E}}^{-1} b_i  \\
- X c_i  &= \Big(\lambda_{i}-{\textstyle \frac{1} {\epsilon_0}} \langle \alpha_{{\scriptscriptstyle E}}^{-1}\rangle  - \gamma_1\Big)b_i
-{\textstyle \frac{1} {\epsilon_0}} \Delta\alpha_{{\scriptscriptstyle E}}^{-1} a_i \\
X b_i  &= \Big(\lambda_{i} - {\textstyle \frac{1}{\epsilon_0}} \langle \alpha_{{\scriptscriptstyle H}}^{-1}\rangle - \gamma_1 \Big)c_i
-{\textstyle \frac{1} {\epsilon_0}}\Delta\alpha_{{\scriptscriptstyle H}}^{-1} d_i \\
- X a_i  &=   \Big(\lambda_{i}-{\textstyle \frac{1} {\epsilon_0}} \langle \alpha_{{\scriptscriptstyle H}}^{-1}\rangle + \gamma_1 \Big)d_i
 - {\textstyle \frac{1} {\epsilon_0}}\Delta\alpha_{{\scriptscriptstyle H}}^{-1} c_i
\end{align}
\end{subequations}
By substituting between these equations, we can obtain the following expressions for the ratios between coefficients:
\begin{petit}
\begin{align}
\frac{a_i}{b_i}  &=
\frac{-(\lambda_{i}+\gamma_1)\frac{1 }{\epsilon_0}\langle\alpha_{{\scriptscriptstyle E}}^{-1}\rangle
-(\lambda_{i}-\gamma_1)\frac{1 }{\epsilon_0}\langle\alpha_{{\scriptscriptstyle H}}^{-1}\rangle
+(\lambda_{i}^2 - \gamma_1^2)
+\frac{1}{2 \epsilon_0^2}(\frac{1}{\alpha^{(1)}_{{\scriptscriptstyle E}}\alpha^{(2)}_{{\scriptscriptstyle H}}}
+ \frac{1}{\alpha^{(2)}_{{\scriptscriptstyle E}}\alpha^{(1)}_{{\scriptscriptstyle H}}})
+X^2}
{(\lambda_{i}+\gamma_1)\frac{1}{\epsilon_0}
(\Delta\alpha_{{\scriptscriptstyle E}}^{-1}- {\Delta\alpha_{{\scriptscriptstyle H}}^{-1}})
+\frac{1}{2 \epsilon_0^2 }(\frac{1}{\alpha^{(2)}_{{\scriptscriptstyle E}}\alpha^{(1)}_{{\scriptscriptstyle H}}}
- \frac{1}{\alpha^{(1)}_{{\scriptscriptstyle E}}\alpha^{(2)}_{{\scriptscriptstyle H}}} )}
\label{eq:eigmodeBegin}
\\[2ex]
\frac{c_i}{d_i}
&=
\frac{(\lambda_{i}-\gamma_1)\frac{1 }{\epsilon_0}\langle\alpha_{{\scriptscriptstyle E}}^{-1}\rangle
+(\lambda_{i}+\gamma_1)\frac{1 }{\epsilon_0}\langle\alpha_{{\scriptscriptstyle H}}^{-1}\rangle
-(\lambda_{i}^2 - \gamma_1^2)
-\frac{1}{2 \epsilon_0^2}(\frac{1}{\alpha^{(1)}_{{\scriptscriptstyle E}}\alpha^{(2)}_{{\scriptscriptstyle H}}}
+ \frac{1}{\alpha^{(2)}_{{\scriptscriptstyle E}}\alpha^{(1)}_{{\scriptscriptstyle H}}})
-X^2}
{(\lambda_{i}+\gamma_1)\frac{1}{\epsilon_0}
(\Delta\alpha_{{\scriptscriptstyle E}}^{-1}- {\Delta\alpha_{{\scriptscriptstyle H}}^{-1}})
+\frac{1}{2 \epsilon_0^2 }(\frac{1}{\alpha^{(2)}_{{\scriptscriptstyle E}}\alpha^{(1)}_{{\scriptscriptstyle H}}}
- \frac{1}{\alpha^{(1)}_{{\scriptscriptstyle E}}\alpha^{(2)}_{{\scriptscriptstyle H}}} )}
\\[2ex]
 \frac{d_i}{b_i} &
 =\frac{1}
{ X \Big(
(\lambda_{i}+\gamma_1)\frac{1}{\epsilon_0}
(\Delta\alpha_{{\scriptscriptstyle E}}^{-1}- {\Delta\alpha_{{\scriptscriptstyle H}}^{-1}})
+\frac{1}{2 \epsilon_0^2 }(\frac{1}{\alpha^{(2)}_{{\scriptscriptstyle E}}\alpha^{(1)}_{{\scriptscriptstyle H}}}
- \frac{1}{\alpha^{(1)}_{{\scriptscriptstyle E}}\alpha^{(2)}_{{\scriptscriptstyle H}}} )
\Big)
}  \; \cdot \nonumber \\
&\qquad \quad \Bigg [ \frac{1}{\epsilon_0}\Big(  \big(\lambda_{i} ^2-{\gamma_1 ^2}\big) \langle\alpha_{{\scriptscriptstyle H}}^{-1}\rangle +
   X^2 \langle\alpha_{{\scriptscriptstyle E}}^{-1}\rangle  \Big)
 - \big(\lambda_{i} +{\gamma_1 }\big)\big(\lambda_{i}^2 - \gamma_1^2 +{X^2}\big)
\nonumber \\ &\qquad \quad  + \big(  \frac{1}{\epsilon_0} \langle\alpha_{{\scriptscriptstyle H}}^{-1}\rangle  -(\lambda_{i} + \gamma_1)\big)
\big(\frac{1}{\alpha^{(1)}_{{\scriptscriptstyle E}} \alpha^{(2)}_{{\scriptscriptstyle E}} \epsilon_0^2}
-\frac{2}{\epsilon _0}{ \langle\alpha_{{\scriptscriptstyle E}}^{-1}\rangle  }\lambda_{i} \big) \Bigg ]\label{eq:eigmodeEnd}
\end{align}
\end{petit}
These three ratios are sufficient to produce expressions for each eigenmode when normalized to $b_i$, a dependence that is removed by simply normalizing each eigenmode.
In Figure~\ref{fig:dimer1}a, we demonstrate that this analytical expression for the eigenmodes is able to accurately predict the optical extinction of an asymmetric dimer made from silicon nanospheres.
The dimer we consider also doubles as an example of how electric to magnetic coupling can be utilized for directional scattering.   
\index{Directional scattering}
The overlap of equal transverse electric and magnetic dipole moments to realize directional scattering through Huygens' condition\cite{Pfeiffer2013}, or weak duality symmetry\cite{Fernandez-CorbatonPRB2013, Zambrana-Puyalto2013},  has largely been obtained by imposing additional geometric freedoms on single nanoparticles.
This has included silicon disks, ellipsoids and other such geometries.  
However, because we only consider the zeroth-order forward and backward scattering, it is not necessary for the electric and magnetic dipole moments to be occupying the same point in space: the transverse displacement of dipoles does not introduce any relative phase shift for this scattering direction.
As such, directional nanoantennas can also be obtained by using {\sl two} distinct nanoparticles where one provides the electric dipole and the other provides the magnetic dipole.  
This is precisely what is prepared in Figure~\ref{fig:comparison}: the magnetic dipole resonance of a 150~nm silicon sphere coincides with the electric resonance of a 200~nm silicon sphere.
When arranging these two spheres as a dimer, the gap between spheres provides control over their coupling strength, which is a freedom we can utilize to equilibrate the electric and magnetic resonances of the single particles.
Indeed, as seen in Figure~\ref{fig:dimer1}b, this approach leads to strong directional scattering as a result of an equality of the total transverse electric and magnetic dipole moments seen in Figure~\ref{fig:dimer1}c.
This has also provided maximum scattering upon directionality, as opposed to similar approaches that produce directional scattering by utilizing Fano resonances, which have a tendency to {\sl reduce} scattering due to eigenmode interference suppressing extinction.

\begin{figure}[!t]
\vspace{5ex}
\centerline{\includegraphics[width=\textwidth]{{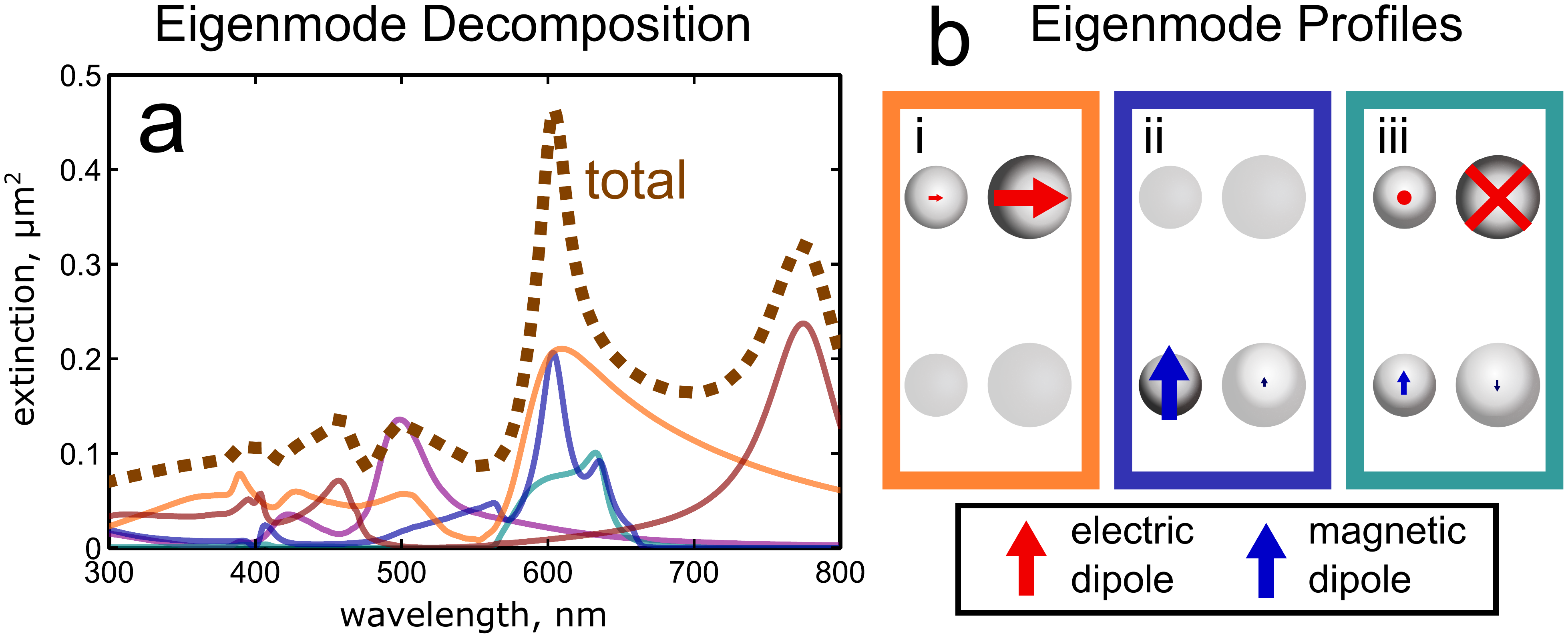}}}
\caption{(a) Eigenmode decomposition 
\index{Eigenmode decomposition}
of the extinction from the silicon nanosphere dimer in Figure~\ref{fig:dimer1}a.  (b) The electric and magnetic dipole moments of the three most dominant eigenmodes that contribute towards the extinction peak at 600~nm. }
\label{fig:dimer2}
\end{figure}
To explore our dimer system further, we can also provide analysis of its behavior from the perspective of the collective eigenmodes.
Figure~\ref{fig:dimer2}a shows the individual eigenmode contributions to extinction, where directional scattering at 600~nm wavelength can be dominantly attributed to the three eigenmodes (i-iii), whose profiles are depicted in Figure~\ref{fig:dimer2}b.
The first two eigenmodes (i and ii) are dominated by the electric and magnetic dipole of the 200~nm and 150~nm nanoparticles, respectively.
A third eigenmode (iii) also contributes to the final magnetic dipole of the 150~nm sphere, although it is dominantly from the electric dipole in the 200~nm sphere aligned parallel to the propagation direction.
The parallel electric dipole does not contribute to forward or backward radiation, but this eigenmode nonetheless provides a second magnetic resonance in the 150~nm sphere, and we can see it compensates the imbalance between the magnitudes of the isolated electric and magnetic dipoles.   
The combination of these three eigenmodes then describes the directional scattering we observed in Figure~\ref{fig:dimer1}b.
It shows that the presented directionality utilizes a triply resonant system to mediate the equal electric and magnetic dipole moments.
This example thereby demonstrates how calculating eigenmodes can provide additional insight into the operation of more complicated dielectric nanoparticle oligomers.

\subsection{Dimensionless Eigenvalues}

The eigenmodes and the eigenvalues we have presented up until now have modeled the polarizability relationship between either dipoles or currents, and the driving field.
However, in some situations, it may also be informative to know the eigenmodes of a geometry where there is no driving field, such as in the transient response to excitation by a pulse of light.
Such eigenmodes, being current distributions $\boldsymbol{\mathcal{J}}_v$, represent the underlying stable currents (or dipole moments) of a given structure: they are the stable oscillations that persist in the given structure when the driving field is removed.
Notably, the associated eigenvalues $\Lambda_v$ are then not polarizabilities, but rather dimensionless values for the self-feedback strength of the given current or dipole distribution.
\begin{align}
\Lambda_v {\boldsymbol{\mathcal{J}}_v(\mathbf{r})}  =&  \; \big( {\bar\epsilon(\mathbf{r})  - \epsilon_0}\big)  \frac{k^2}{\epsilon_0}  \int _{V}  {\bar{G_0}}(\mathbf{r},\mathbf{r}') \boldsymbol{\mathcal{J}}_v(\mathbf{r}') \;{\mathrm{d}^3x'}
\label{eq:different eig}
\end{align}
Comparing this expression to Equation~\ref{eq:eigenmode equation}, we can see that there is no distinction between eigenmodes with or without a driving field when the given structure is made up of a homogeneous isotropic material.
It is only when we introduce additional materials into our scattering system that the stable resonances are changed by the presence of a driving field.
However, if we consider the analogous situation for the dipole model in Equation~\ref{eq:ED model}, the eigenmodes of Equations~\ref{eq:eigenmode equation} and \ref{eq:different eig} become distinct whenever there is more than one single isotropic polarizability describing all dipoles.
So we should expect a distinction between the two types of eigenmodes whenever a dipole system consists of both electric and magnetic dipoles.
Moreover, we can consider eigenmodes $\{\boldsymbol{\scriptstyle \mathcal{P}}_v,\,\boldsymbol{\scriptstyle \mathcal{M}}_v\}$ of the electric and magnetic dipole model in Equation~\ref{eq:Dmodel} when there is no driving field.
\begin{subequations}
\begin{align}
\Lambda_v\boldsymbol{\scriptstyle \mathcal{P}}_{v}^{(i)}  = &
\alpha_{E}^{(i)}k^{2}
  \bigg(\underset{j\neq i}{{\sum}}\bar{G}_{0}(\mathbf{r}_{i},\mathbf{r}_{j})\cdot  \boldsymbol{\scriptstyle \mathcal{P}}_{v}^{(j)}
-\frac{1}{c_0}\,\nabla\times\bar{G}_{0}(\mathbf{r}_{i},\mathbf{r}_{j}) \cdot \boldsymbol{\scriptstyle \mathcal{M}}_{v}^{(j)}\bigg)  \\
\Lambda_v \boldsymbol{\scriptstyle \mathcal{M}}_{v}^{(i)} =  &
 \alpha_{H}^{(i)}k^{2}  \bigg(\underset{j\neq i}{{\sum}}\bar{G}_{0}(\mathbf{r}_{i},\mathbf{r}_{j}) \cdot \boldsymbol{\scriptstyle \mathcal{M}}_{v}^{(j)}
+c_0\,\nabla\times\bar{G}_{0}(\mathbf{r}_{i},\mathbf{r}_{j}) \cdot \boldsymbol{\scriptstyle \mathcal{P}}_{v}^{(j)}\bigg)
\end{align}
\end{subequations}
This expression will produce different eigenmodes to that of Equation~\ref{eq:DmodelEig}, but it instead provides different insights into the given scattering system.
Notably, we do not need to scale the electric and magnetic dipoles to have correct units for the eigenvalue: it now remains unitless for any scaling between electric and magnetic dipoles.

\section{Observation of Fano Resonances in Nanoparticle Oligomers}
\label{sec:3}
 \index{Oligomers}
The study of Fano resonances 
\index{Fano resonance}
in plasmonic nanoparticle structures has more than a decade of history, whereas
the theoretical predictions~\cite{Miroshnichenko2012,HopkinsPoddubny2013} and experimental
demonstrations~\cite{Chong2014,Filonov2012,HopkinsFilonovPRB2015,HopkinsFilonov2015,Yan2015}
of Fano resonances in all-dielectric nanoparticle oligomers 
\index{Dielectric nanoparticles}
have been reported only recently.  
In comparison to the oligomers composed of regular plasmonic nanoparticles, all-dielectric oligomer structures
support both strong electric and magnetic resonances
\index{Magnetic resonance}
 in their individual constituent nanoparticles, opening a new avenue for mutual inter-element coupling and, consequently,
for the formation of novel collective resonances. 
\index{Optical magnetism}
The existence of Fano resonances in silicon oligomers also suggested promise for nonlinear applications, owing largely to silicon's higher resistance to heat compared to the metals in plasmonic nanostructures, and combined with a decent capacity for optically induced free carrier generation.
\index{Nanophotonics}

\begin{figure}[!ht]
\centering
\includegraphics[width=0.9\textwidth]{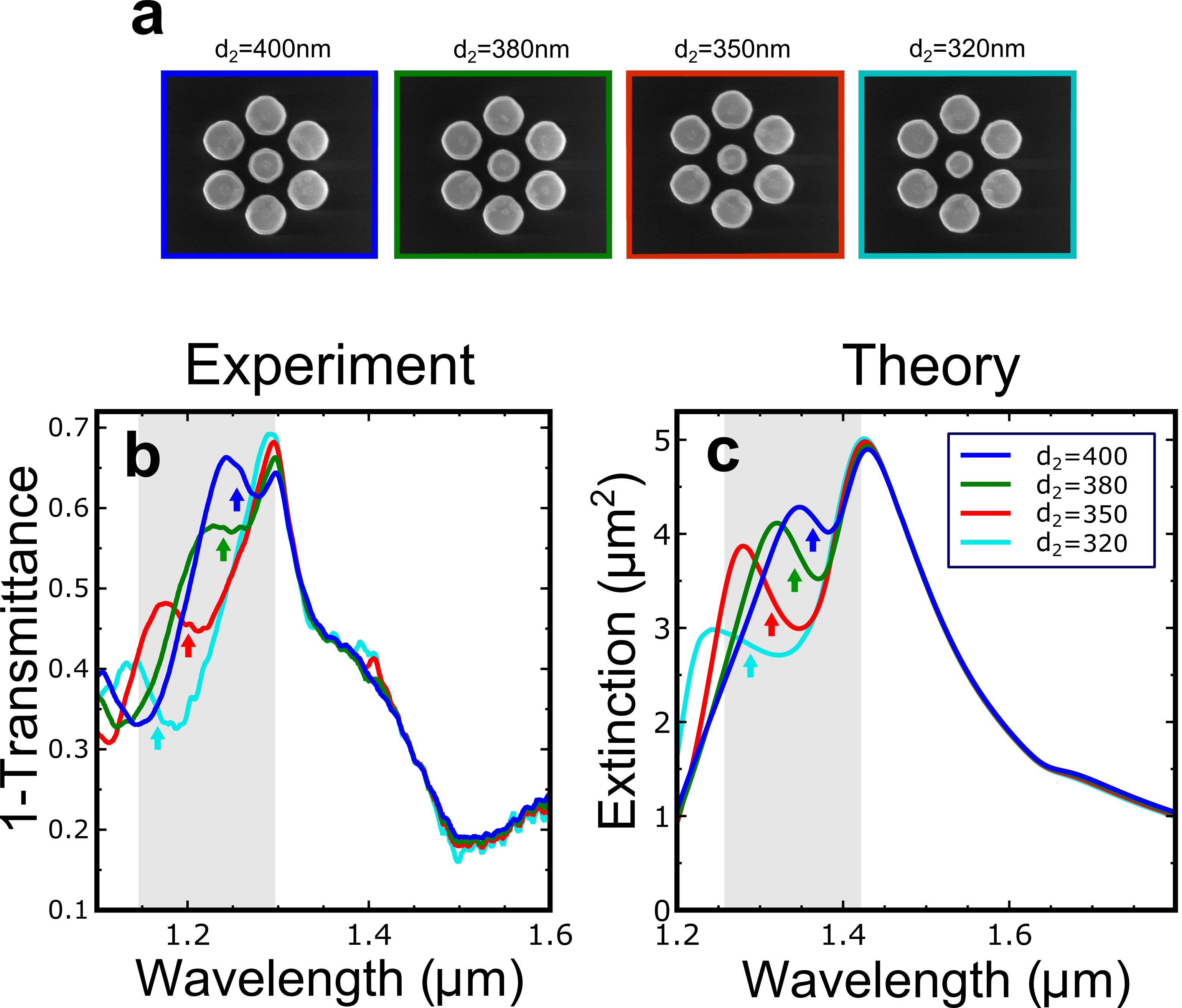}
\caption{ The observation of Fano resonances from silicon oligomers~\cite{Chong2014}.
(a) SEM images of the considered heptamer oligomers made from silicon
nanodisks with variations of central nanodisk diameter, and the corresponding (b) experimentally measured transmission spectra and (c) simulated extinction spectra. 
A Fano resonance is created in the heptamers
(indicated by colored arrows in the gray-shaded region), and shifts across
the spectrum as the diameter of the central particle is varied.}
\label{fig:1}
\end{figure}

For such context, the optical Fano resonance is usually recognized as arising from the interference of spectrally broad and narrow resonances.
This largely owes to the original asymmetric line-shape in the energy spectrum of atomic photoionization explained by Fano~\cite{Fano1961} to be due to constructive and destructive two-channel  interference between photoionization from a broad ground state and from a discrete autoionized state of an atom.  
However, in nanophotonic systems, interference also regularly occurs between resonances of comparable spectral width~\cite{HopkinsPoddubny2013}.
The resulting asymmetric line-shape can still be clearly observed in both theory and experiment since the destructive and constructive interference occurs over a narrow spectral range with strongly resonant response.
Owing to these shared characteristics, such features have also been branded as Fano resonances.
To this extent, Figure~\ref{fig:1} presents experimental results for the observation of such Fano resonances
in  heptamer oligomers composed of silicon nanodisks.  
Figjre~\ref{fig:1}a shows SEM images of four heptamer structures, which are each arranged in two-dimensional arrays
with a lattice constant of 2840 nm. 
The spacing between heptamers was chosen to ensure that the dominant coupling remains {\sl within} each individual heptamer, while keeping the surface density of these oligomer structures sufficiently high to create pronounced resonances in tranmission.
The measurements of optical transmission from different silicon nanodisk heptamer arrays, with varying central nanodisk diameter and unpolarized plane wave illumination, are presented in Figure~\ref{fig:1}b.
The experimental and theoretical extinction spectra show qualitative agreement. 
By computing the collective eigenmodes of the oligomer structures, corresponding to the dominant resonances, the mechanism of eigenmode interference was able to be seen in the heptamer structures~\cite{Chong2014}. 
The peak at longer wavelengths corresponds to the magnetic resonance of the collective structure, the peak at shorter wavelengths  is associated with the magnetic resonance of the central particle.
This confirmed that the observed Fano resonances were indeed originating from interference between the optically-induced magnetic resonances 
\index{Magnetic resonance}
associated with the central particles and those of the surrounding structure.

Filonov {\it et al.}~\cite{Filonov2014} later observed the existence of Fano resonances in all-dielectric oligomer structures with a series of microwave experiments.
In these experiments, they were able to directly confirm the role of the central particle by using far-field and near-field measurements to observe both the forward scattering and the local phase distribution of the magnetic near-field.
The resonant suppression of scattering was observed to be accompanied by a $\pi$~phase difference of the near-field at the central particle, relative to that of the surrounding ring of particles. 
This was supported by analysis of the expected eigenmodes when using the coupled-dipole approximation, and allowed clear identification of the collective resonances that were contributing to the observed interference phenomena. 

\begin{figure}[!ht]
\centering
\includegraphics[width=0.95\textwidth]{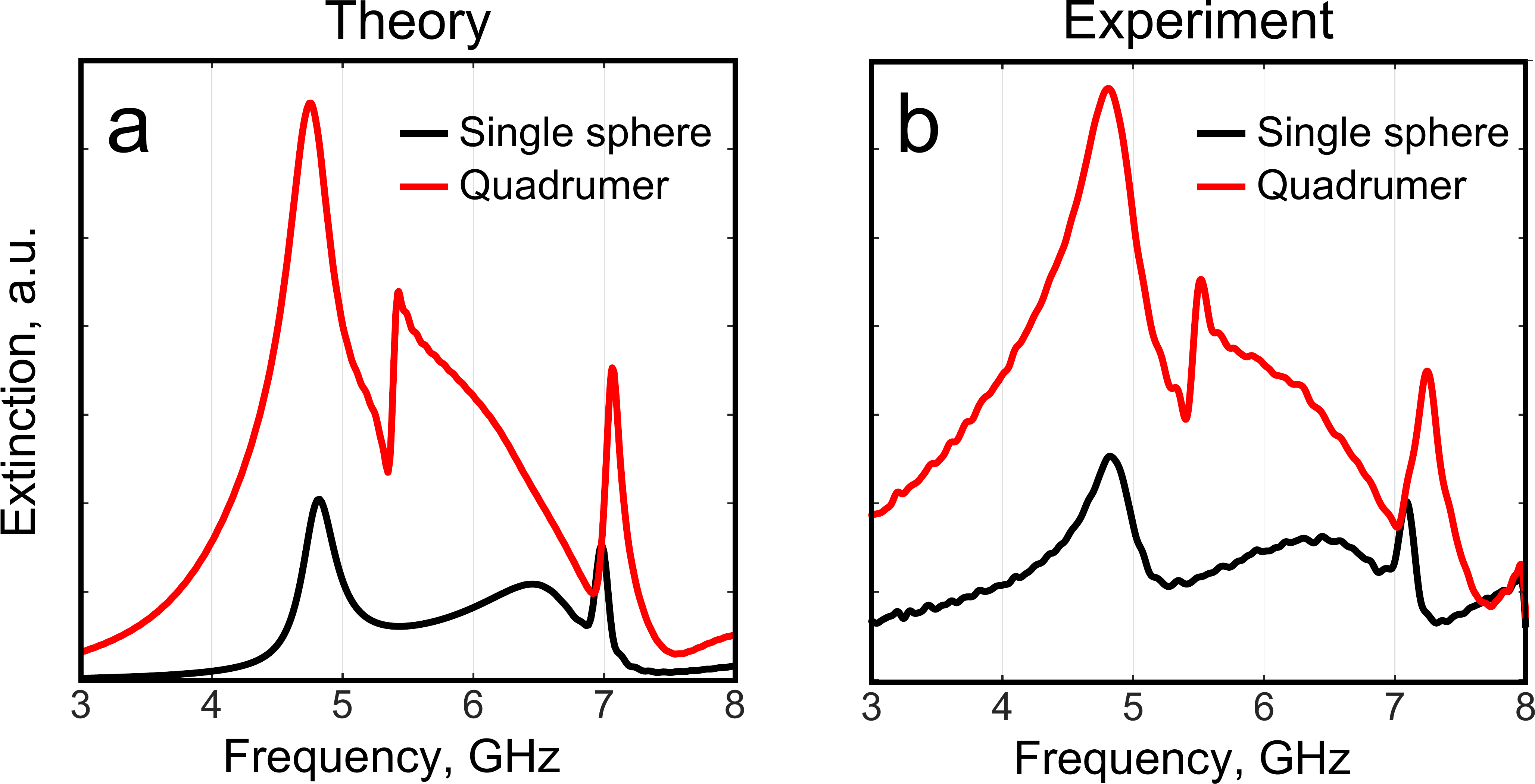}
\caption{Fano resonances in the transmission spectrum of a dielectric nanoparticle quadrumer due to coupling between collective and individual magnetic dipoles~\cite{HopkinsFilonov2015}.
Shown is (a) numerical simulation and (b) experimental
measurement of extinction for a quadrumer made of four \mbox{MgO-TiO$_2$}
ceramic spheres ($\epsilon=16$) under oblique plane wave illumination.  
The extinction spectrum for a single
\mbox{MgO-TiO$_2$} sphere is shown for reference. 
A sharp Fano resonance is seen at 5.4
GHz, which emerges from interference between two collective magnetic dipole-like resonances, {\it cf.}~\cite{HopkinsFilonov2015}.}
\label{fig:2}
\end{figure}

Given dielectric nanoparticles allow additional coupling mechanisms between electric and magnetic resonances, one should also expect Fano resonances to emerge for even a small number of dielectric particles such as trimers~\cite{HopkinsFilonovPRB2015} and quadrumers~\cite{HopkinsFilonov2015}.
In the latter work, a magnetic field polarized along a symmetric quadrumer's principal axis of rotation
was observed to induce a resonant collective circulation of electric field, while also coupling with the inherent magnetic responses of the individual nanoparticles. 
Both such collective responses appear as magnetic dipoles in the far-field, and their interference can be tailored to produce distinctive and sharp {\sl magnetic} Fano resonances~\cite{HopkinsFilonov2015}, {\it i.e.} originating due to induced magnetic--magnetic coupling.
\index{Magnetic Fano resonance} 
These theoretical predictions were also explored experimentally~\cite{HopkinsFilonov2015}, by studying the scattering properties of \mbox{MgO-TiO$_2$} ceramic spheres with a dielectric permittivity of 16 at 9--12~GHz. 
These ceramic spheres in the microwave range therefore have very similar properties to silicon nanospheres in
the optical range, and can be employed as a useful macroscopic analogue to prototype silicon nanostructures.
The experimentally measured, and numerically calculated, scattered power from the quadrumer structure is shown in Figure~\ref{fig:2}. 
It can clearly be seen that a magnetic Fano resonance is produced at 5.4~GHz in both simulation and experiment. 
This was the first example of a magnetic--magnetic Fano resonance in a single symmetric metamolecule. 
\index{Optical magnetism}
Notably, this Fano resonance occurs near the intersection of the single particle's magnetic resonance (lowest frequency) and electric resonance (second lowest frequency), reflecting that the collective resonances we consider are dependent on both the electric and the magnetic dipole polarizabilities of individual spheres.
The role of individual electric and magnetic dipoles has also been considered relating to Fano resonances arising from the interaction between broad and narrow modes in dimers composed of two different silicon nanospheres~\cite{Yan2015}, which can be used to suppress backward scattering at Fano wavelengths. \index{Directional scattering}  \index{Dimer}
This directional scattering was much more prominent than that from a single silicon sphere and shows promising applications in areas such as directional nanoantennas or optical switching, opening up future avenues for developing all-dielectric nanophotonic devices.

\section{Concluding Remarks}

The study of resonant dielectric and semiconductor nanoparticle structures has become a new research direction in modern nanophotonics, where they are expected to complement or substitute a number of existing plasmonic components.
This can be largely attributed to the intrinsic propensity of high-index dielectric nanoparticles for Mie-type electric and magnetic resonances, combined with significantly lower Ohmic losses.  
Here we have shown that all-dielectric nanostructures are indeed able to reproduce a range of subwavelength resonant effects that were previously considered as realizable only in plasmonic  nanostructures. 
The subsequent co-existence of strong electric and magnetic resonances, their interference, and the resonant enhancement of magnetic fields, all bring new physics and novel functionalities even to simple dielectric nanoparticle oligomer geometries.  
These developments now underlie a host of promise for future low-loss nanophotonic devices, and especially in the nonlinear regime.

\bibliographystyle{ieeetr}
\bibliography{bibliography}

\begin{thebibliography}{10}

\bibitem{Evlyukhin2010}
A.~B. Evlyukhin, C.~Reinhardt, A.~Seidel, B.~S. Luk'yanchuk, and B.~N.
  Chichkov, ``{Optical response features of Si-nanoparticle arrays},'' {\em
  Phys. Rev. B}, vol.~82, p.~045404, 2010.

\bibitem{Krasnok2011_Jetp}
A.~E. Krasnok, A.~E. Miroshnichenko, P.~A. Belov, and Y.~S. Kivshar, ``{Huygens
  optical elements and Yagi-Uda nanoantennas based on dielectric
  nanoparticles},'' {\em JETP. Lett.}, vol.~94, no.~8, p.~593, 2011.

\bibitem{GarciaEtxarri2011}
A.~Garc\'ia-Etxarri, R.~G\'omez-Medina, L.~S. Froufe-P\'erez, C.~L\'opez,
  L.~Chantada, F.~Scheffold, J.~Aizpurua, M.~Nieto-Vesperinas, and J.~J.
  S\'aenz, ``Strong magnetic response of submicron silicon particles in the
  infrared,'' {\em Opt. Express}, vol.~19, pp.~4815--4826, 2011.

\bibitem{Kerker1975}
M.~Kerker, ``Invisible bodies,'' {\em J. Opt. Soc. Am.}, vol.~65, pp.~376--379,
  1975.

\bibitem{Nieto2011}
M.~Nieto-Vesperinas, R.~Gomez-Medina, and J.~J. Saenz, ``Angle-suppressed
  scattering and optical forces on submicrometer dielectric particles,'' {\em
  J. Opt. Soc. Am. A}, vol.~28, no.~1, pp.~54--60, 2011.

\bibitem{Rolly2012}
B.~Rolly, B.~Stout, and N.~Bonod, ``Boosting the directivity of optical
  antennas with magnetic and electric dipolar resonant particles,'' {\em Opt.
  Express}, vol.~20, no.~18, pp.~20376--20386, 2012.

\bibitem{Geffrin2012}
J.~M. Geffrin, B.~García-C\'amara, P.~A. R.~G\'omez-Medina, L.~S.
  Froufe-P\'erez, C.~Eyraud, A.~Litman, R.~Vaillon, F.~Gonz\'ale\'z,
  M.~Nieto-Vesperinas, J.~J. S\'aenz, and F.~Moreno, ``Magnetic and electric
  coherence in forward- and back-scattered electromagnetic waves by a single
  dielectric subwavelength sphere,'' {\em Nat. Commun.}, vol.~3, p.~1171, 2012.

\bibitem{Kuznetsov2012}
A.~I. Kuznetsov, A.~E. Miroshnichenko, Y.~H. Fu, J.~B. Zhang, and
  B.~Luk'yanchuk, ``Magnetic light,'' {\em Sci. Rep.}, vol.~2, p.~492, 2012.

\bibitem{Evlyukhin2012NL}
A.~B. Evlyukhin, S.~M. Novikov, U.~Zywietz, R.~L. Eriksen, C.~Reinhardt, S.~I.
  Bozhevolnyi, and B.~N. Chichkov, ``Demonstration of magnetic dipole
  resonances of dielectric nanospheres in the visible region,'' {\em Nano
  Lett.}, vol.~12, pp.~3749--3755, 2012.

\bibitem{Liu2012_ACSNANO}
W.~Liu, A.~E. Miroshnichenko, D.~N. Neshev, and Y.~S. Kivshar, ``Broadband
  unidirectional scattering by magneto-electric core-shell nanoparticles,''
  {\em ACS Nano}, vol.~6, no.~6, p.~5489, 2012.

\bibitem{Fu2013}
Y.~H. Fu, A.~I. Kuznetsov, A.~E. Miroshnichenko, Y.~F. Yu, and B.~Luk'yanchuk,
  ``Directional visible light scattering by silicon nanoparticles,'' {\em Nat.
  Commun.}, vol.~4, p.~1527, 2013.

\bibitem{Liu2014_CPB}
W.~Liu, A.~E. Miroshnichenko, and Y.~S. Kivshar, ``Control of light scattering
  by nanoparticles with optically-induced magnetic responses,'' {\em Chin.
  Phys. B}, vol.~23, no.~23, p.~047806, 2014.

\bibitem{Krasnok2012_OE}
A.~E. Krasnok, A.~E. Miroshnichenko, P.~A. Belov, and Y.~S. Kivshar,
  ``All-dielectric optical nanoantennas,'' {\em Opt. Express}, vol.~20, no.~18,
  p.~20599, 2012.

\bibitem{Filonov2012}
D.~S. Filonov, A.~E. Krasnok, A.~P. Slobozhanyuk, P.~V. Kapitanova, E.~A.
  Nenasheva, Y.~S. Kivshar, and P.~A. Belov, ``Experimental verification of the
  concept of all-dielectric nanoantennas,'' {\em Appl. Phys. Lett.}, vol.~100,
  p.~201113, 2012.

\bibitem{Krasnok204_nanoscale}
A.~Krasnok, C.~Simovski, P.~Belov, and Y.~S.~Kivshar, ``Superdirective
  dielectric nanoantenna,'' {\em Nanoscale}, vol.~6, pp.~7354--7361, 2014.

\bibitem{Miroshnichenko2010}
A.~E. Miroshnichenko, S.~Flach, and Y.~S. Kivshar, ``Fano resonances in
  nanoscale structures,'' {\em Rev. Mod. Phys.}, vol.~82, pp.~2257--2298, 2010.

\bibitem{Lukyanchuk2010}
B.~Luk'yanchuk, N.~I. Zheludev, S.~A. Maier, N.~J. Halas, P.~Nordlander,
  H.~Giessen, and C.~T. Chong, ``{The Fano resonance in plasmonic
  nanostructures and metamaterials},'' {\em Nature Materials}, vol.~9,
  pp.~707--715, 2010.

\bibitem{Joe2006}
Y.~S. Joe, A.~M. Satanin, and C.~S. Kim, ``Classical analogy of Fano
  resonances,'' {\em Phys. Scr.}, vol.~74, pp.~259--266, 2006.

\bibitem{Gallinet2011}
B.~Gallinet and O.~J.~F. Martin, ``Ab initio theory of Fano resonances in
  plasmonic nanostructures and metamaterials,'' {\em Phys. Rev. B}, vol.~83,
  p.~235427, 2011.

\bibitem{Miroshnichenko2012}
A.~E. Miroshnichenko and Y.~S. Kivshar, ``Fano resonances in all-dielectric
  oligomers,'' {\em Nano Lett.}, vol.~12, pp.~6459--6463, 2012.

\bibitem{Sheinfux2014}
H.~H. Sheinfux, I.~Kaminer, Y.~Plotnik, G.~Bartal, and M.~Segev,
  ``Subwavelength multilayer dielectrics: Ultrasensitive transmission and
  breakdown of effective-medium theory,'' {\em Phys. Rev. Lett.}, vol.~113,
  p.~243901, 2014.

\bibitem{HopkinsPoddubny2013}
B.~Hopkins, A.~N. Poddubny, A.~E. Miroshnichenko, and Y.~S. Kivshar,
  ``{Revisiting the physics of Fano resonances for nanoparticle oligomers},''
  {\em Phys. Rev. A}, vol.~88, p.~053819, 2013.

\bibitem{KuznetsovSPIE2015}
A.~I. Kuznetsov, ``Light manipulation by resonant dielectric nanostructures and
  metasurfaces,'' in {\em Proc. SPIE 9544, Metamaterials, Metadevices, and
  Metasystems 2015}, p.~95442A, 2015.

\bibitem{Staude2013}
I.~Staude, A.~E. Miroshnichenko, M.~Decker, N.~T. Fofang, S.~Liu, E.~Gonzales,
  J.~Dominguez, T.~S. Luk, D.~N. Neshev, I.~Brener, and Y.~Kivshar, ``Tailoring
  directional scattering through magnetic and electric resonances in
  subwavelength silicon nanodisks,'' {\em ACS Nano}, vol.~7, no.~9,
  pp.~7824--7832, 2013.

\bibitem{Lukiyanchuk2015}
B.~S. Luk'yanchuk, N.~V. Voshchinnikov, R.~Paniagua-Dom{\'\i}nguez, and A.~I.
  Kuznetsov, ``Optimum forward light scattering by spherical and spheroidal
  dielectric nanoparticles with high refractive index,'' {\em ACS Photon.},
  vol.~2, no.~7, pp.~993--999, 2015.

\bibitem{Chong2014}
K.~E. Chong, B.~Hopkins, I.~Staude, A.~E. Miroshnichenko, J.~Dominguez,
  M.~Decker, D.~N. Neshev, I.~Brener, and Y.~S. Kivshar, ``{Observation of Fano
  Resonances in All-Dielectric Nanoparticle Oligomers},'' {\em Small}, vol.~10,
  no.~10, pp.~1985--1990, 2014.

\bibitem{HopkinsFilonov2015}
B.~Hopkins, D.~S. Filonov, A.~E. Miroshnichenko, F.~Monticone, A.~Al\`u, and
  Y.~S. Kivshar, ``Interplay of magnetic responses in all-dielectric oligomers
  to realize magnetic Fano resonances,'' {\em ACS Photon.}, vol.~2,
  pp.~724--729, 2015.

\bibitem{Pfeiffer2013}
C.~Pfeiffer and A.~Grbic, ``Metamaterial huygens' surfaces: Tailoring wave
  fronts with reflectionless sheets,'' {\em Phys. Rev. Lett.}, vol.~110,
  p.~197401, May 2013.

\bibitem{Decker2015}
M.~Decker, I.~Staude, M.~Falkner, J.~Dominguez, D.~N. Neshev, I.~Brener,
  T.~Pertsch, and Y.~S. Kivshar, ``{High-Efficiency Dielectric Huygens'
  Surfaces},'' {\em Adv. Opt. Mater.}, vol.~3, pp.~813--820, 2015.

\bibitem{Chong2015}
K.~E. Chong, I.~Staude, A.~James, J.~Dominguez, S.~Liu, S.~Campione, G.~S.
  Subramania, T.~S. Luk, M.~Decker, I.~B. Dragomir N. Neshev~and, and Y.~S.
  Kivshar, ``Polarization-independent silicon metadevices for efficient optical
  wavefront control,'' {\em Nano Lett.}, vol.~15, no.~8, pp.~5369--5374, 2015.

\bibitem{Arbabi2015}
A.~Arbabi, Y.~Horie, M.~Bagheri, and A.~Faraon, ``Dielectric metasurfaces for
  complete control of phase and polarization with subwavelength spatial
  resolution and high transmission,'' {\em Nat. Mat.}, vol.~10, pp.~937--944,
  2015.

\bibitem{Yaghjian1980}
A.~D. Yaghjian, ``{Electric Dyadic Green`s Functions in the Source Region},''
  {\em Proc. IEEE}, vol.~68, no.~2, pp.~248--263, 1980.

\bibitem{Powell2014}
D.~A. Powell, ``Resonant dynamics of arbitrarily shaped meta-atoms,'' {\em
  Phys. Rev. B}, vol.~90, p.~075108, 2014.

\bibitem{Hentschel2010}
M.~Hentschel, M.~Saliba, R.~Vogelgesang, H.~Giessen, A.~P. Alivisatos, and
  N.~Liu, ``Transition from isolated to collective modes in plasmonic
  oligomers,'' {\em Nano Lett.}, vol.~10, pp.~2721--2726, 2010.

\bibitem{HopkinsLiu2013}
B.~Hopkins, W.~Liu, A.~E. Miroshnichenko, and Y.~S. Kivshar, ``Optically
  isotropic responses induced by discrete rotational symmetry of nanoparticle
  clusters,'' {\em Nanoscale}, vol.~5, pp.~6395--6403, 2013.

\bibitem{Rahmani2013}
M.~Rahmani, E.~Yoxall, B.~Hopkins, Y.~Sonnefraud, Y.~Kivshar, M.~Hong,
  C.~Phillips, S.~A. Maier, and A.~E. Miroshnichenko, ``Plasmonic nanoclusters
  with rotational symmetry: Polarization-invariant far-field response vs
  changing near-field distribution,'' {\em ACS Nano}, vol.~7, pp.~11138--11146,
  2013.

\bibitem{Mulholland1994}
G.~W. Mulholland, C.~F. Bohren, and K.~A. Fuller, ``Light scattering by
  agglomerates: Coupled electric and magnetic dipole method,'' {\em Langmuir},
  vol.~10, no.~8, p.~2533, 1994.

\bibitem{Draine1994}
B.~T. Draine and P.~J. Flatau, ``Discrete-dipole approximation for scattering
  calculations,'' {\em J. Opt. Soc. Am. A}, vol.~11, pp.~1491--1499, Apr 1994.

\bibitem{Mie1908}
G.~Mie, ``Beitrage zur optik truber medien,'' {\em Ann. Phys.}, vol.~25,
  pp.~377--445, 1908.

\bibitem{Bohren1983}
C.~F. Bohren and D.~R. Huffman, {\em Absorption and scattering of light by
  small particles}.
\newblock New York: Wiley, 1983.

\bibitem{Chen2011}
J.~Chen, J.~Ng, Z.~Lin, and C.~T. Chan, ``Optical pulling force,'' {\em Nat.
  Photon.}, vol.~5, pp.~531--534, 2011.

\bibitem{Grahn2012}
P.~Grahn, A.~Shevchenko, and M.~Kaivola, ``Electromagnetic multipole theory for
  optical nanomaterials,'' {\em New J. Phys.}, vol.~14, p.~093033, 2012.

\bibitem{Miroshnichenko2015}
A.~E. Miroshnichenko, A.~B. Evlyukhin, Y.~F. Yu, R.~M. Bakker, A.~Chipouline,
  A.~I. Kuznetsov, B.~Luk'yanchuk, B.~N. Chichkov, and Y.~S. Kivshar,
  ``Nonradiating anapole modes in dielectric nanoparticles,'' {\em Nat.
  Commun.}, vol.~6, p.~8069, 2015.

\bibitem{Arango2013}
F.~B. Arango and A.~F. Koenderink, ``Polarizability tensor retrieval for
  magnetic and plasmonic antenna design,'' {\em New J. Phys.}, vol.~15,
  p.~073023, 2013.

\bibitem{LandauLifshitzVol5}
L.~D. Landau, E.~M. Lifshitz, and L.~P. Pitaevski{\u\i}, {\em Statistical
  Physics, Part 1}, vol.~5 of {\em Course of theoretical physics}.
\newblock Pergamon Press Ltd., 3~ed., 1980.

\bibitem{Gantmacher1959}
F.~Gantmacher, {\em The Theory of Matrices}.
\newblock Chelsea Publishing Company, 1959.

\bibitem{Craven1969}
B.~Craven, ``Complex symmetric matrices,'' {\em J. Austral. Math. Soc.},
  vol.~10, pp.~341--354, 1969.

\bibitem{Merchiers2007}
O.~Merchiers, F.~Moreno, F.~Gonzalez, and J.~M. Saiz, ``Light scattering by an
  ensemble of interacting dipolar particles with both electric and magnetic
  polarizabilities,'' {\em Phys. Rev. A.}, vol.~76, no.~4, p.~043834, 2007.

\bibitem{Kern2010}
A.~M. Kern and O.~J.~F. Martin, ``Pitfalls in the determination of optical
  cross sections from surface integral equation simulations,'' {\em IEEE Trans.
  Ant. Prop.}, vol.~58, no.~6, pp.~2158--2161, 2010.

\bibitem{HopkinsPoddubny2016}
B.~Hopkins, A.~N. Poddubny, A.~E. Miroshnichenko, and Y.~S. Kivshar,
  ``{Circular dichroism induced by Fano resonances in planar chiral
  oligomers},'' {\em Laser Photon. Rev.}, vol.~10, no.~1, pp.~137--146, 2016.

\bibitem{Forestiere2013}
C.~Forestiere, L.~D. Negro, and G.~Miano, ``{Theory of coupled plasmon modes
  and Fano-like resonances in subwavelength metal structures},'' {\em Phys.
  Rev. B}, vol.~88, p.~155411, 2013.

\bibitem{Frimmer2012}
M.~Frimmer, T.~Coenen, and A.~F. Koenderink, ``{Signature of a Fano Resonance
  in a Plasmonic Metamolecule's Local Density of Optical States},'' {\em Phys.
  Rev. Lett.}, vol.~108, p.~077404, 2012.

\bibitem{Lovera2013}
A.~Lovera, B.~Gallinet, P.~Nordlander, and O.~J. Martin, ``{Mechanisms of Fano
  Resonances in Coupled Plasmonic Systems},'' {\em ACS Nano}, vol.~7 (5),
  pp.~4527--4536, 2013.

\bibitem{HopkinsFilonovPRB2015}
B.~Hopkins, D.~S. Filonov, S.~B. Glybovski, and A.~E. Miroshnichenko,
  ``{Hybridization and the origin of Fano resonances in symmetric nanoparticle
  trimers},'' {\em Phys. Rev. B}, vol.~92, p.~045433, 2015.

\bibitem{Heiss2001}
W.~D. Heiss, ``Exceptional points of non-hermitian operators,'' {\em J. Phys.
  A: Math Gen.}, vol.~37, p.~2455, 2001.

\bibitem{Dembowski2001}
C.~Dembowski, H.-D. Graf, H.~L. Harney, A.~Heine, W.~D. Heiss, H.~Rehfeld, and
  A.~Richter, ``The physics of exceptional points,'' {\em Phys. Rev. Lett.},
  vol.~86, pp.~787--790, 2001.

\bibitem{Heiss2012}
W.~D. Heiss, ``The physics of exceptional points,'' {\em J. Phys. A: Math.
  Theor.}, vol.~45, p.~444016, 2001.

\bibitem{Yan2015}
J.~Yan, P.~Liu, Z.~Lin, H.~Wang, H.~Chen, C.~Wang, and G.~Yang, ``{Directional
  Fano resonances in a silicon nanoparticle dimer},'' {\em ACS Nano}, vol.~9,
  pp.~2968--2980, 2015.

\bibitem{Bakker2015}
R.~M. Bakker, D.~Permyakov, Y.~F. Yu, D.~Markovich, R.~Paniagua-Dom{\'\i}nguez,
  L.~Gonzaga, A.~Samusev, Y.~Kivshar, B.~Luk'yanchuk, and A.~I. Kuznetsov,
  ``Magnetic and electric hotspots with silicon nanodimers,'' {\em Nano Lett.},
  vol.~15, no.~3, pp.~2137--2142, 2015.

\bibitem{Zywietz2015}
U.~Zywietz, M.~K. Schmidt, A.~B. Evlyukhin, C.~Reinhardt, J.~Aizpurua, and
  B.~N. Chichkov, ``Electromagnetic resonances of silicon nanoparticle dimers
  in the visible,'' {\em ACS Photon.}, vol.~2, pp.~913--920, 2015.

\bibitem{Miroshnichenko2011}
A.~E. Miroshnichenko, B.~Luk'yanchuk, S.~A. Maier, and Y.~S. Kivshar,
  ``Optically induced interaction of magnetic moments in hybrid
  metamaterials,'' {\em ACS Nano}, vol.~6, no.~1, pp.~837--842, 2011.

\bibitem{Fernandez-CorbatonPRB2013}
I.~Fernandez-Corbaton and G.~Molina-Terriza, ``Role of duality symmetry in
  transformation optics,'' {\em Phys. Rev. B}, vol.~88, p.~085111, 2013.

\bibitem{Zambrana-Puyalto2013}
X.~Zambrana-Puyalto, I.~Fernandez-Corbaton, M.~L. Juan, X.~Vidal, and
  G.~Molina-Terriza, ``{Duality symmetry and Kerker conditions},'' {\em Opt.
  Lett.}, vol.~38, no.~11, pp.~1857--1859, 2013.

\bibitem{Fano1961}
U.~Fano, ``Effects of configuration interaction on intensities and phase
  shifts,'' {\em Phys. Rev.}, vol.~124, pp.~1866--1878, 1961.

\bibitem{Filonov2014}
D.~S. Filonov, A.~P. Slobozhanyuk, A.~E. Krasnok, P.~A. Belov, E.~A. Nenasheva,
  B.~Hopkins, A.~E. Miroshnichenko, and Y.~S. Kivshar, ``{Near-field mapping of
  Fano resonances in all-dielectric oligomers},'' {\em Appl. Phys. Lett.},
  vol.~104, p.~021104, 2014.

\end{thebibliography}

\printindex 
\end{document}